\documentclass{aa}  
\usepackage{graphicx}
\usepackage[dvipsnames]{xcolor}
\usepackage{hyperref}
\hypersetup{colorlinks=true,linkcolor=blue,citecolor=blue,filecolor=blue,urlcolor=blue}
\usepackage{txfonts}
\DeclareRobustCommand{\ion}[2]{\textup{#1\,\textsc{\lowercase{#2}}}}

\newcommand{\kms}{\ensuremath{{\rm km~s^{-1}}}}

\newcommand{\beix}{$^{9}$Be\,}

\newcommand{\mygi}{MyGIsFOS}

\begin{document} 

   \title{A  Sequoia stellar candidate  with very high $^7$Li and $^9$Be}

 \titlerunning{A Sequoia candidate with very high $^7$Li and $^9$Be}

  \author{L. Monaco\inst{1,2}
          \and E. Caffau \inst{3,2}
          \and P. Molaro\inst{2,4}
          \and P. Bonifacio\inst{3,2}
          \and G. Cescutti \inst{5,2,6,4}
}
          
\institute{Universidad Andres Bello, Facultad de Ciencias Exactas, Departamento de F{\'i}sica y Astronom{\'i}a - Instituto de Astrof{\'i}sica, Autopista
Concepci\'on-Talcahuano 7100, Talcahuano, Chile\\
\email{lorenzo.monaco@unab.cl}
\and
INAF-OATs, Via G.B.Tiepolo 11, Trieste, I 34143, Italy
\and
LIRA, Observatoire de Paris, Universit{\'e} PSL,
Sorbonne Universit{\'e}, Universit{\'e} Paris Cit{\'e}, CY Cergy Paris Universit{\'e},
CNRS,92190 Meudon, France
\and
Institute  of Fundamental Physics of the Universe, IFPU, Via Beirut, 2, Trieste, I-34151, Italy
\and
Dipartimento di Fisica, Universit{\`a} degli Studi di Trieste, via Tiepolo 11, I-34143 Trieste, Italy
\and 
INFN, Sezione di Trieste, Via A. Valerio 2, 34127 Trieste, Italy
             }


 
\abstract
{}
{The metal-poor star BPM\,3066 belongs to the retrograde halo and presents unexpectedly strong spectral features of lithium. To gain insight into the origin of this peculiar abundance, we investigate the chemistry and kinematic properties of this star.}
{ We performed a local thermodynamic equilibrium chemical abundance analysis of 
 UVES/VLT high-resolution spectra of BPM\,3066 using ATLAS9 and ATLAS12 model atmospheres and the MyGIsFOS code. We further characterised the orbital properties of the star by integrating its orbit and analysing its integrals of motion using the galpy code.}
{The star BPM\,3066 shows an exceptional overabundance of both lithium and beryllium. The abundances are A(Li) = 3.0 and A(Be) = 2.1, which are respectively about  0.8 and 2.2 dex higher than the Li and Be abundances expected at [Fe/H] = -1.5, the  metallicity of the star. The observed ratio 
$^7$Li/$^9$Be 
is 7.9, which is  close to that   expected from a synthesis by spallation processes. Overabundances of  Si, Al,  and of the neutron capture elements Sr,Y, Zr, and Ba are also  measured. Kinematically, BPM\,3066  has an eccentric, strongly retrograde orbit, confined to a height lower than 1\,kpc from the galactic plane, and  it is a candidate member of the Sequoia/Thamnos accreted galaxy.}
{The processes leading to the $^7$Li and $^9$Be synthesis could have occurred in the environment of a hypernova. This is supported by some abundance anomalies like the high value of Si, [Si/Fe]=1.2 and [Si/O]=1.1. However,  the simultaneous high values of N, Na, Al, Sc, Ti, and Cu are at odds with the expectations from a hypernova. Alternatively, the abundances of BPM\,3066 could result from the engulfing of rocky planets that were rich in spallated Li and Be. 
In both cases, it is remarkable that such an extreme abundance 
pattern has been found in a star belonging to the Sequoia/Thamnos accreted galaxy.}

   \keywords{
   Nuclear reactions, nucleosynthesis, abundances - Galaxy: stellar content - Galaxy:structure - Stars: abundances - Stars: Population II - Planetary systems
               }

   \maketitle
%

\section{Introduction}

The origin of the light elements lithium and beryllium is different from that of the other elements that are  burnt and synthesised in the hot interior of  stars. Li has  multiple origins as it is made in the Big Bang, in the interstellar medium (ISM) by spallation, in asymptotic giant branch  stars (AGBs), 
and in novae \citep[][]{molaro23}. On the other hand, $^{9}$Be is a pure product of spallation processes of 
CNO nuclei by energetic cosmic rays or of energetic CNO nuclei 
onto protons and $\alpha$ particles at rest in the ISM \citep[][]{reeves70}.

Unevolved halo stars show a flat plateau in the A(Li) versus [Fe/H] plane, the so-called Spite plateau \citep[][A(Li)=2.2]{Spite1982}, which bears witness to the primordial production of lithium, even though the Li abundances observed in old, unevolved metal-poor stars are significantly smaller than expected from primordial nucleosynthesis when the baryon-to-photon  density is estimated by the primordial deuterium abundance \citep[][A(Li)=2.72]{coc17}. 

In contrast to Li, no significant $^{9}$Be (the only stable Be isotope) production is expected in standard big bang nucleosynthesis, and a steady, roughly linear increase in the $^{9}$Be abundance with metallicity is observed, which suggests that Be behaves like a primary element \citep[][]{molaro1997A&A...319..593M,Smiljanic2009,Boesgaard2011}, and that there is an enrichment mechanism acting on a galactic scale. Several authors   have investigated a possible flattening of this relation at the lowest  metallicities, which might indicate the presence of a plateau pointing to a primordial component suggested by inhomogeneous cosmologies \citep{malaney1989, primas2000A&A...362..666P,Smiljanic21}.
There is an intrinsic scatter in the Be-Fe relation  that could be attributed to the contribution of the different galactic sub-components that were assembled at early epochs \citep[][]{Smiljanic21}, since the cosmic ray flux and CNO abundances could have been different between galaxies, and  a different growth rate could be possible. \citet{molaro2020MNRAS.496.2902M} show that the [Be/H] versus [Fe/H] relation of candidate stars of Gaia-Sausage-Enceladus \citep[][GSE, the latest major accretion event in the Milky Way]{Belokurov2018,Haywood2018,Helmi2018} is less scattered and has a different gradient in comparison  to the Milky Way.

We report here on the detection of a metal-poor, lithium-rich dwarf star that simultaneously, presents an exceptionally large Be abundance and an overall peculiar chemical composition. We discuss the possible origin of this star and its peculiarities. 

The paper is organised as follows: in Sect.\,\ref{sec:obs}, we report on the observations. In Sect. \ref{sec:abu} we describe the chemical abundance analysis, and in Sect. \ref{sec:kin} we provide a kinematic characterisation of the star. We note that the star turns out to be a candidate Sequoia (or Thamnos) star \citep[][]{barba19,myeong19,villanova19,koppelman19}, another of the Galactic halo sub-structures detected in the Gaia era.  In Sect.\,\ref{sec:dis} we discuss our results, including the possible origin of the observed chemical patterns. Finally, we summarise our findings in Sect. \ref{sec:sum}.

\section{Target and observations}\label{sec:obs}

Star BPM\,3066 was first listed in the Bruce Proper Motion Survey as possessing an appreciable proper motion\footnote{'A Catalogue of 904 Stars in the Southern Hemisphere with Proper Motions Exceeding 0T5 Annually,' Pub. Minnesota Obs., 3, No. 1, 1941}. The survey was carried out in the 1930s by Willem Luyten using the  24-inch Harvard Observatory Bruce Telescope after it had been moved to South Africa and  preliminary photometric colours had been measured for the star \citep{luyten1942ApJ....96...55L}. The star was later observed as part of the GALactic Archaeology with HERMES (GALAH) survey \citep[][]{buder2021MNRAS.506..150B} and \citet[][]{simpson2021MNRAS.507...43S} identified the star as a lithium-rich dwarf belonging to the retrograde halo (see  Sect.\,\ref{sec:kin}).

The Gaia $G$ magnitude of BPM\,3066 is 12.78. Observations of BPM\,3066, also known as Gaia DR3 4667364088963367808, GALAH 131116000501386, and 2MASS J03355522-6833454, have been obtained with the Ultraviolet and Visual Echelle Spectrograph \citep[UVES,][]{dekker00} at the European Southern Observatory Very Large Telescope (ESO-VLT) under program 111.24HT.001, P.I L. Monaco. A single 593-second exposure was acquired at airmass 1.5 on the night between the 17$^{}$ and the 18$^{}$  July, 2023. We adopted the DIC1 346+580 UVES setting, which covers the 304.4-391.7\,nm and 472.6-683.5\,nm spectral ranges with the blue and red arm, respectively. We further adopted two by two on chip binning (50 kHz readout speed, high gain). The blue arm spectrum includes the $^9$BeII $\lambda\lambda$ 313.0422, 313.1067\,nm
resonance doublet, while the red arm  covers the Li resonance doublet at 670.78\,nm. We adopted a slit width of 1.2$^{\prime\prime}$ in
the blue arm and 1.0$^{\prime\prime}$ in the red arm, 
which deliver a spectral resolving power of 37,000 and 42,000, respectively.
We retrieved the reduced data\footnote{https://www.eso.org/sci/software/pipelines/} from the ESO archive.\footnote{http://archive.eso.org/wdb/wdb/adp/phase3\_main/form}

The signal-to-noise (S/N) ratio is $\simeq 100$ at 671\,nm and $\simeq 15$ at 313\,nm, respectively.
The stellar radial velocity  has been measured using the red arm spectrum through a cross-correlation function \citep[][]{tonry79} with a synthetic spectrum calculated with atmospheric parameters similar to that of the star using the {\tt fxcor} task within IRAF v2.18\footnote{https://iraf.noirlab.edu/release} \citep[][]{tody86,tody93,fitzpatrick24}.
From the UVES spectra, we obtained  a radial velocity of 263.16 $\pm 0.13$ km s$^{-1}$ , which is in agreement with the value of 263.02 $\pm 2.53$ of Gaia and the 264.04 $\pm 0.25$ km s$^{-1}$ of GALAH, indicating that BPM\,3066 is not a radial velocity variable.

\section{Chemical abundance analysis}\label{sec:abu}

The stellar atmospheric parameters were obtained with the iterative procedure described in 
\citet[][see also \citealt{bonifacio24}]{lombardo21}. Briefly, we adopted the reddening of the star from the \citet[][]{Schlafly2011} maps. The de-reddened Gaia G$_{BP}$-G$_{RP}$ colour was compared with synthetic photometry in order to derive the stellar effective temperature, $\rm T_{eff}$ , and the bolometric correction. Then, the surface gravity, $\rm \log{g,}$ was obtained from the Stefan-Boltzmann equation. Finally, the extinction coefficients were updated and the whole procedure was iterated until convergence ($\rm \Delta T_{eff}<50$\,K and $\rm\Delta\log{g}<0.05$\,dex). The synthetic colours were computed from the Mucciarelli et al. (in preparation) grid of ATLAS9 fluxes.  

We used the parameters obtained in this way to derive the stellar metallicity using the MyGIsFOS code \citep[][]{sbordone14} and, with this new metallicity, we repeated the procedure once more to obtain the final stellar parameters. The microturbulence was derived using the \citet[][]{mashonkina17} calibration. 
The grid of synthetic spectra employed were the same described in \citet{lombardo2023} and were computed with SYNTHE \citep{Kurucz2005} 
using a grid of ATLAS 12 \citep{Kurucz2005} model atmospheres (Sbordone, in preparation). 
We performed an additional run to derive the chemical abundances using the MyGIsFOS code with the final stellar parameters. The abundances obtained were used to compute an ATLAS12 model atmosphere \citep[][]{Kurucz2005}, 
which was used to compute a dedicated grid of synthetic spectra in which the model atmosphere is
always the same and all abundances are scaled, at steps of 0.2\,dex between --1 and +1 with respect
to those of the starting model. The synthetic spectra  in this
case were computed with {\tt turbospectrum} \citep{alvarezplez,plez2012}.
This non-standard use of \mygi\ has the advantage of circumventing the issue
with ionised elements highlighted by \citet{bonifacio24}, as demonstrated in
\citet{caffau_sdss_2024}.
From a \mygi\ run with this dedicated grid, we obtained the final chemical abundances, which are reported in Table\,\ref{tab:abboq053}. 

For BPM\,3066, we adopted $\rm T_{eff}=5910$\,K, $\rm\log{g}=4.29$\,dex, and $\xi=1.15$\,\kms. The iron metallicity obtained from 196 neutral iron lines is $\rm [Fe/H] = -1.57$\,dex, while the value from 20 singly  ionised iron lines is $\rm [Fe/H] = -1.52$\,dex.

The GALAH survey adopted very similar parameters ($\rm T_{eff}=5874$\,K, $\rm\log{g}=4.29$\,dex, and $\xi =1.19$\,\kms), but obtained a metallicity about 0.2\,dex larger ($\rm [Fe/H]=-1.33$\,dex, \citealt{buder2021MNRAS.506..150B}). 
The GALAH collaboration only provides  non-local thermodynamic equilibrium (NLTE) abundances, thus their abundances are not
directly comparable to ours. 
For Fe, NLTE corrections are based on the computations of \citet{amarsi2016}, who adopt the theory
by \citet{barklem2016} to describe the collisions with H atoms. 
Our estimate of the NLTE correction for this star is +0.04\,dex
\citep{bergemann2012}, which is based on the assumption that the collision with neutral hydrogen can be
represented by the `Drawin' formalism \citep{drawin1968,drawin1969,SH1984}, assuming $S_H=1,$
and
provides NLTE corrections that are systematically smaller than those
of \citet{amarsi2016}. For example, for star HD\,140283, \citet{amarsi2016} provide an Fe abundance
that is 0.1\,dex higher than that of \citet{bergemann2012}.
Thus, at least 0.14\,dex of the difference between our Fe and that of GALAH can be ascribed
to the NLTE corrections adopted by GALAH; the rest of the difference is well within the error
of either result. For BPM\,3066 the Gaia DR3 astrophysical parameters catalogue \citep[][]{gaiadr3} reports  photometric temperatures, surface gravities, and metallicities of teff\_gspphot=6082\,K, logg\_gspphot=4.35\,dex, and mh\_gspphot=--1.43, which are in remarkably good agreement with ours.

\begin{figure}
 \includegraphics[width=0.5\textwidth]{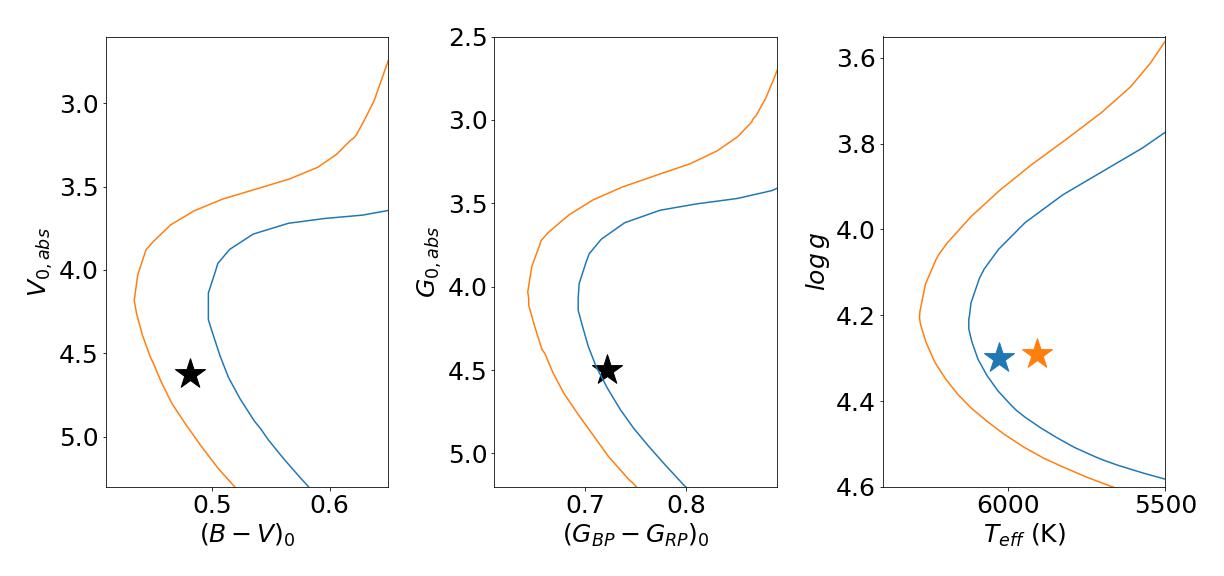}
    \caption{Left and middle panels: V {\it vs} (B-V) and Gaia G {\it vs} ($G_{BP}-G_{RP}$) colour-magnitude diagrams (CMDs). Right panel: Kiel diagram, log\,g {\it vs} $\rm T_{eff}$. Star BPM\,3066 is indicated by the filled star. In the right panel, the orange star is for the parameters we adopt here, while the blue one is for the GALAH BSTEP parameters. In all panels, two PARSEC isochrones of metallicity and ages ([M/H], Age)=(-0.71, 11.7) and (-1.26, 13.0) are shown in blue and orange, respectively.}
    \label{fig:cmd}
\end{figure}

In Fig.\,\ref{fig:cmd}, we show the position of BPM\,3066 (filled star) in the absolute optical UCAC4 V versus (B-V) \citep[left panel][]{ucac4} and Gaia G versus ($G_{BP}-G_{RP}$) (middle panel) colour-magnitude diagrams (CMDs). The right panel shows the position of BPM\,3066 in the Kiel diagram. The orange star is for the $\rm T_{eff}$, log\,g parameters we adopt, while the blue one is for the parameters estimated by the BSTEP code \citep[`Bayesian Stellar Parameter Estimation',][]{sharma18} as reported in the corresponding GALAH DR3 value-added catalogue, namely $\rm T_{eff}$=6031\,K, log\,g=4.3. These values were obtained by combining a set of observables with PARSEC isochrones \citep[`PAdova and TRieste Stellar Evolution Code',][]{bressan12}. In all panels, we also plot two PARSEC isochrones of metallicity and age (Gyr) ([M/H], Age)=(-0.71, 11.7) (blue isochrone) and (-1.26, 13.0) (orange isochrone). The former are GALAH DR3 BSTEP values. The latter is the global metallicity we obtained from the \citet[][]{salaris05} formula, using the iron abundance we measured and an $\alpha$ enhancement of [$\alpha$/Fe]=+0.4. We also adopted a reference age of 13 Gyr. 

\begin{figure}
 \includegraphics[width=0.5\textwidth]{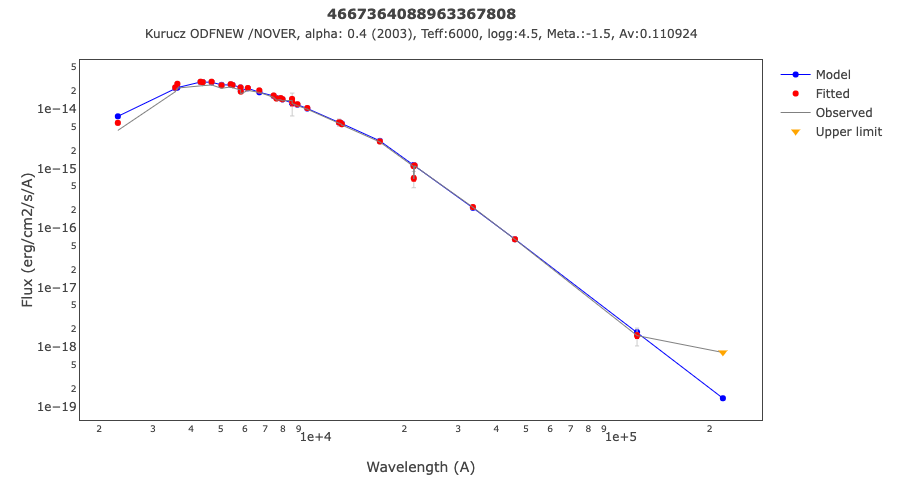}
    \caption{Spectral energy distribution of BPM\,3066  as obtained using the VOSA service \citep[][]{vosa08}. We also show a fit with an ATLAS 9 model flux with parameters similar to those derived here.}
    \label{fig:sed}
\end{figure}

The position of BPM\,3066 in all diagrams clearly indicates that the star is currently on the main sequence, even though the orange isochrone does not provide a perfect fit to BPM\,3066, particularly in the Kiel diagram, where some ambiguity between the main sequence and sub-giant branch remains. A gravity 0.45 (0.23) smaller (larger) would be required for the stars to fit the sub-giant/main-sequence sections of the orange isochrone.The blue isochrone provides a good fit to the position of BPM\,3066  on the Gaia CMD and in the Kiel diagram for the BSTEP parameters, as expected. However, the surface effective BSTEP metallicity ([M/H]=-0.71$\pm$0.14) is larger than that obtained from the GALAH DR3 spectroscopic iron abundance ([Fe/H]=-1.33) and $\alpha$ enhancement ([$\alpha$/Fe]=+0.54), namely [M/H]=-0.90, a value intermediate between the two plotted isochrones. We notice that the BSTEP atmospheric parameters shown in the right panel are not those recommended for use by the GALAH collaboration, which, as already mentioned, are very similar to our values. 
Finally, the stellar spectral energy distribution of BPM\,3066 is fitted properly by an ATLAS 9 model flux with parameters similar to those adopted here, from the near ultraviolet GALEX (Galaxy Evolution Explorer) band to the infrared WISE (Wide-field Infrared Survey Explore) W3 band (see Fig.\,\ref{fig:sed}).

\subsection{Light elements Li and Be}
\begin{figure}
 \includegraphics[width=0.5\textwidth]{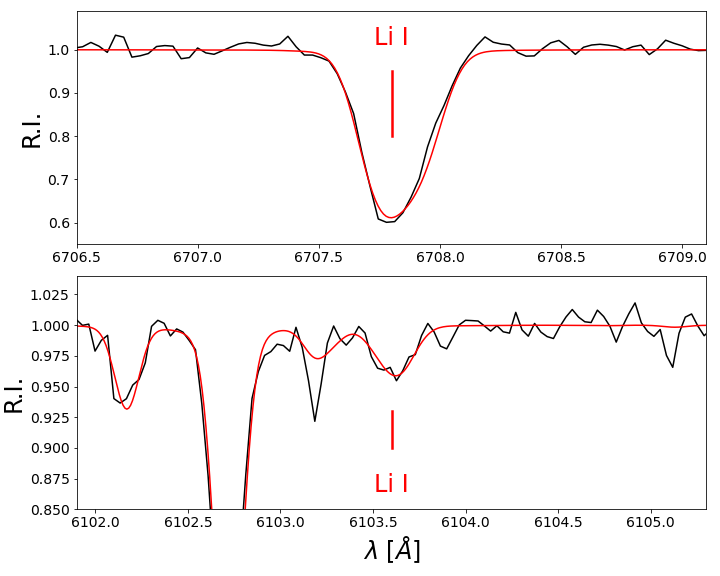}
    \caption{Observed spectrum (solid black) compared to the best fit (solid red) with A(Li) = 3.0. The UVES \ion{Li}{I}  670.78 nm region of BPM\,3066 is shown in the top panel. The  \ion{Li}{I}  $\lambda\lambda$ 610.36 nm 2$^2$P-3$^2$D subordinate transition is shown in the bottom panel; the strong line on the blue side is the   \ion{Ca}{I} 610.2723 nm line.}
    \label{fig:fig1a}
\end{figure}

The  UVES spectrum of BPM\,3066 around the \ion{$^7$Li}{I}  region is shown in the upper panel of Fig.\,\ref{fig:fig1a} where the observed spectrum is  compared to the best fit  of the \ion{$^7$Li}{i} 670.78 nm resonance doublet. In the bottom panel of Fig.\,\ref{fig:fig1a}, the best fit of the \ion{$^7$Li}{i}  $\lambda\lambda$ 610.36 nm 2$^2$P-3$^2$D subordinate transition is also shown. 
The lower level of the \ion{$^7$Li}{i}  610.36 line is the upper level of the  670.78 
2$^2$S-2$^2$P transition, which samples different portions of the stellar atmosphere.
An average $^7$Li abundance of A($^7$Li) = 3.0 is derived from the line-profile fit of the two Li lines, at 610.36\,nm and 670.78\,nm. We note, however, that in order to satisfactorily reproduce the profile of the resonance doublet at 670.78 nm, no $^6$Li needed to be adopted. The GALAH survey reports a lithium abundance of A(Li)=3.02 for this star, which is very close to our value.

\begin{figure}
 \includegraphics[width=0.45\textwidth]{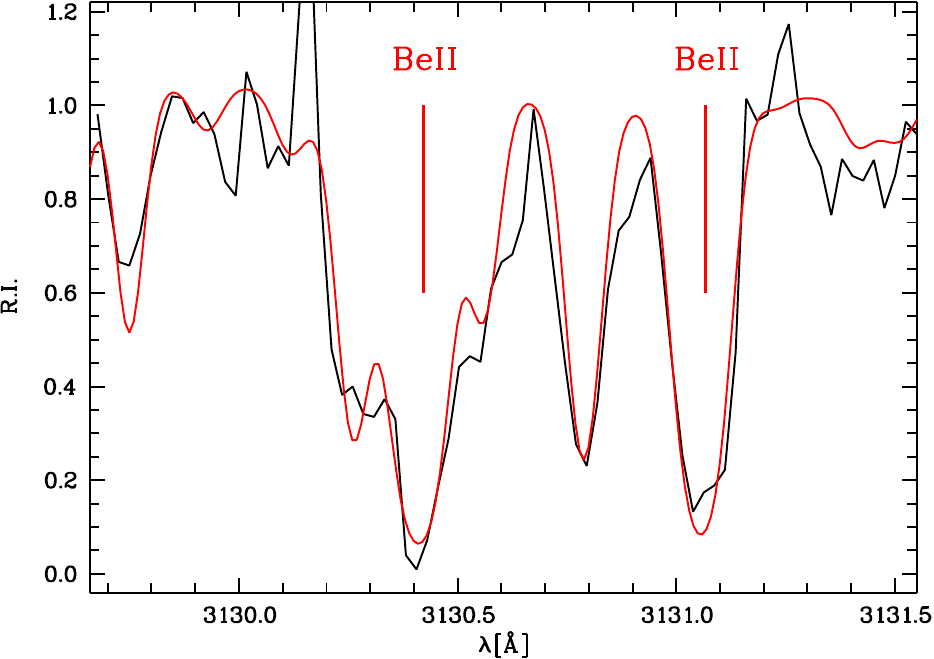}
    \caption{UVES spectrum of BPM\,3066 around the $^9$\ion{Be}{II}  $\lambda\lambda$ 313.0422, 313.1067  nm resonance doublet  region. The observed spectrum (solid black) is compared to the synthesis (solid red) on both \ion{Be}{ii} 313.0422 and 313.1067 nm  resonance lines, with  an abundance of A($^9$Be) = 2.10, the best fit value of the redder line.
    The $^9$Be region is a complex  one  and the stronger line of the \ion{Be}{II}  doublet is blended with  \ion{V}{II} 313.0269,  \ion{Fe}{II} 313.0565,   \ion{Ti}{II} 313.0798 nm, and several other weaker    lines. }
    \label{fig:fig2}
\end{figure}

The  UVES spectrum of BPM\,3066 around the \ion{$^9$Be}{II} region is shown in Fig. \ref{fig:fig2}. The Be region is a complex one  and the stronger line of the \ion{$^9$Be}{II} doublet at 313.0 nm is blended with the \ion{V}{II} and \ion{Fe}{I} lines. We adopt a $^9$Be abundance of A($^9$Be) = 2.1, which is the best-fit abundance to the isolated line at 313.1\,nm, but which also reproduces well the complex region at 313.04\,nm. We note that the corresponding \ion{$^7$Be}{II} lines (313.0583\,nm, 313.1228\,nm) would be distinguished from the \ion{$^9$Be}{II} ones (313.0422\,nm, 313.1067\,nm) at the resolution of the UVES spectrum (separation $\Delta v = 15.4\,\kms$). However, given the short half-life of $^7$Be (53.22 days), as expected, lines from this isotope are not detected. We confirmed the reliability of the reduction process in the spectral region of the Be lines by performing a manual data reduction using the UVES pipeline, and we obtained practically indistinguishable results compared with the reduced spectrum retrieved from the ESO archive.

\begin{figure}
 \includegraphics[width=0.45\textwidth]{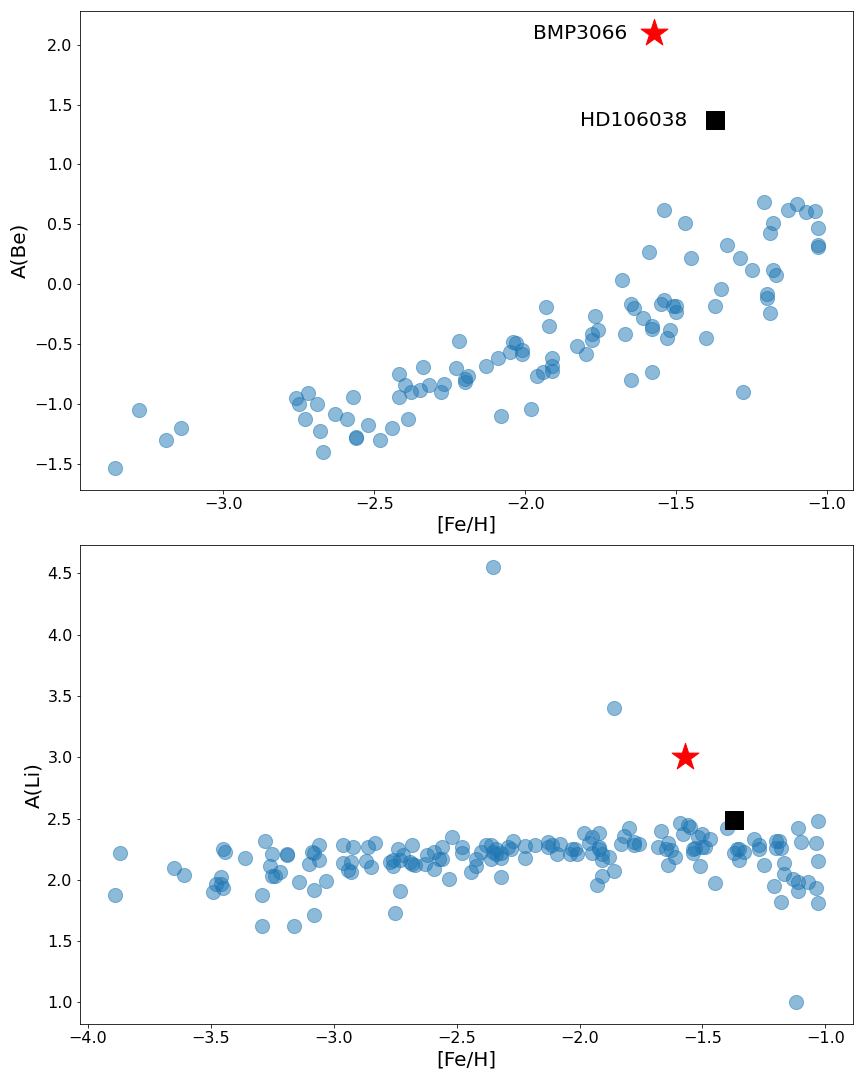}
    \caption{Abundances of A(Be) (upper panel) and A(Li) (bottom panel) in  BPM\,3066 (red star) as a function of [Fe/H]. Blue circles are stars with $\rm 5700<T_{eff}<7000$\,K, $\rm\log{g}>3.65$, and  $\rm [Fe/H]<-1$\,dex, (without upper limits) as retrieved from the SAGA database \citep[][]{SAGA}. Star HD106038 is marked with the black filled square, where the iron and lithium abundances are from \citet[][]{Asplund2006} and the beryllium abundance is from \citet[][]{tan2009}. The SAGA [Fe/H] abundances have been modified to account for the different reference solar iron abundance adopted here.}
    \label{fig:fig3}
\end{figure}

Both the Li  (see bottom panel of Fig.\,\ref{fig:fig3}) and the Be (upper panel) abundances of BPM\,3066 (red star) are much higher than expected for the metallicity of the star (blue filled circles). Besides BPM\,3066, the two stars with A(Li)$>$3.0 in the lower panel of Fig.\,\ref{fig:fig3} are LAMOSTJ074102.07+213246.6 and LAMOSTJ075816.39+470343.3 from \citet[][]{li2018ApJ...852L..31L}. We plan future  investigations also of their Be content.

\subsection{Other elements}

Star BPM\,3066  looks clearly enhanced in both Li and Be, but also shows peculiarities in other elements.
We fitted the NH band at 335\,nm and we derived a low N abundance, $\rm [N/H]=-1.76$\,dex (see Table\,\ref{tab:abboq053}).

We derived all the other elements using MyGIsFOS. In this way we derived abundances for 20 additional elements, reported in Table\,\ref{tab:abboq053}.
Oxygen was derived from OH molecules, while the other abundances are based on atomic lines.
In Table\,\ref{tab:abboq053}, we report the abundances (A(X), [X/H]), the standard deviation ($\sigma$), [X/Fe] (with Fe from neutral lines in the case of neutral species and from ionised Fe lines in the case of ionised species), and the NLTE correction with the corresponding reference.
The lines used are provided in electronic form. For elements measured from one feature only, we indicate a conservative uncertainty of 0.2\,dex. In the case of Be, in the table we report the difference between the value measured from the two features but, as already mentioned, we adopt the abundance measured from the redder one.

The star is enhanced in $\alpha$ elements (Mg, Ca, Ti), but it is rich in Si ($\rm [Si/Fe]=+1.17$\,dex from neutral lines and $+1.32$\,dex from ionised lines).
The star is also rich in Al: the two \ion{Al}{i} lines provide an [Al/Fe] ratio of almost +1\,dex. The general chemical pattern is rather peculiar, with a remarkable overabundance of Si, Al, and of the neutron capture elements Sr, Y, Zr, and Ba.

\begin{table*}
\caption{Local thermodynamic equilibrium abundances of BPM\,3066.} 
\label{tab:abboq053}
\begin{tabular}{lrrrrrrrrrl}
\hline
\smallskip
 El & Z  &ion & A(X)$_\sun$ &Nlines& A(X) &    [X/H] & $\sigma$  & [X/Fe] & NLTE & Ref. \\
\hline
 Li &  3 & 0 &   3.28    &  2   & 3.00 &        &  0.09 &       \\
 Be &  4 & 1 &   1.38    &  2   & 2.10 &        &  0.20 &        \\
 NH &  7 &  0 & 7.86 & band &6.10 & $-1.76$ &  0.20 & $-0.19$ \\
 OH &  8 &  0 & 8.76 & 40   &7.49 & $-1.27$ &  0.31 & $ 0.30$ \\
 Na & 11 &  0 & 6.30 &  3   &4.92 & $-1.38$ &  0.16 & $ 0.19$ \\
 Mg & 12 &  0 & 7.54 &  2   &6.59 & $-0.95$ &  0.20 & $ 0.62$ & $0.02$ & \citet{bergemann2017} \\ 
 Al & 13 &  0 & 6.47 &  2   &5.89 & $-0.58$ &  0.02 & $ 0.99$ \\
 Si & 14 &  0 & 7.52 &  8   &7.12 & $-0.40$ &  0.03 & $ 1.17$ & $-0.01$ & \citet{bergemann2013} \\ 
 Si & 14 &  1 & 7.52 &  3   &7.32 & $-0.20$ &  0.07 & $ 1.32$ \\
 Ca & 20 &  0 & 6.33 & 19   &5.15 & $-1.18$ &  0.06 & $ 0.39$ & $0.05$  & \citet{mashonkina2007} \\ 
 Sc & 21 &  0 & 3.10 &  1   &2.00 & $-1.10$ &  0.20 & $ 0.47$ \\
 Sc & 21 &  1 & 3.10 &  8   &2.09 & $-1.01$ &  0.11 & $ 0.51$ \\
 Ti & 22 &  0 & 4.90 & 13   &3.69 & $-1.21$ &  0.07 & $ 0.36$ & $0.31$  & \citet{bergemann2011}  \\
 Ti & 22 &  1 & 4.90 & 28   &3.73 & $-1.17$ &  0.15 & $ 0.35$ & $0.01$  &  \citet{bergemann2011} \\
 V  & 23 &  0 & 4.00 &  1   &3.05 & $-0.95$ &  0.20 & $ 0.62$ \\
 V  & 23 &  1 & 4.00 &  6   &2.67 & $-1.33$ &  0.22 & $ 0.19$ \\
 Cr & 24 &  0 & 5.64 &  7   &4.09 & $-1.55$ &  0.12 & $ 0.02$ & $0.27$   & \citet{bergemann2010} \\ 
 Cr & 24 &  1 & 5.64 &  8   &4.33 & $-1.31$ &  0.06 & $ 0.21$ \\
 Mn & 25 &  0 & 5.37 &  6   &3.47 & $-1.90$ &  0.17 & $-0.33$ \\
 Mn & 25 &  1 & 5.37 &  3   &3.71 & $-1.66$ &  0.01 & $-0.14$ \\
 Fe & 26 &  0 & 7.52 &196   &5.95 & $-1.57$ &  0.12 &         & $0.04$   & \citet{bergemann2012} \\ 
 Fe & 26 &  1 & 7.52 & 20   &6.00 & $-1.52$ &  0.11 &         \\
 Co & 27 &  0 & 4.92 & 17   &3.33 & $-1.59$ &  0.15 & $-0.02$ & $0.45$  & \citet{bergemannco2010} \\ 
 Ni & 28 &  0 & 6.23 & 21   &4.79 & $-1.44$ &  0.27 & $ 0.13$ \\
 Ni & 28 &  1 & 6.23 &  1   &4.98 & $-1.25$ &  0.20 & $ 0.27$ \\
 Cu & 29 &  0 & 4.21 &  3   &2.64 & $-1.57$ &  0.03 & $ 0.00$ \\
 Zn & 30 &  0 & 4.62 &  2   &3.22 & $-1.40$ &  0.07 & $ 0.17$ \\
 Sr & 38 &  1 & 2.92 &  2   &2.05 & $-0.87$ &  0.05 & $ 0.65$ \\
 Y  & 39 &  1 & 2.21 &  7   &1.38 & $-0.83$ &  0.10 & $ 0.69$ \\
 Zr & 40 &  1 & 2.62 &  4   &1.88 & $-0.74$ &  0.09 & $ 0.78$ \\
 Ba & 56 &  1 & 2.17 &  1   &1.70 & $-0.47$ &  0.20 & $ 1.05$ \\
\hline
\end{tabular}
\tablefoot{
The adopted solar abundances are from \citet[][]{Lodders2009}, except for N, O, and Fe, which are from \citet[][]{caffau11b}, and Zn, which is from \citet[][]{caffau11a}. Li is from the Orgueil CI-chondrite \citep[][]{Lodders2009}.
}
\end{table*}

\section{Kinematics}\label{sec:kin}
Stars belonging to different populations show widely differing
kinematical behaviour. Some components of the Milky Way are rapidly rotating with little dispersion in the velocities of the members, while some others are slowly rotating but with a high dispersion.

\begin{figure}
 \includegraphics[width=0.5\textwidth]{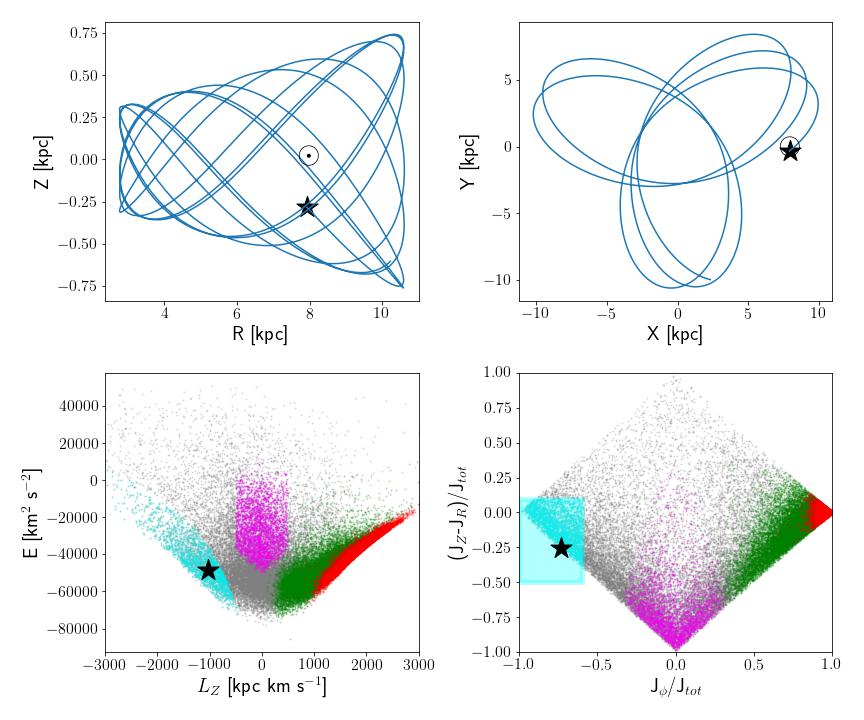}
    \caption{Upper panels: Orbit of BPM\,3066 in the plane of the Galaxy (Y {\it vs} X, right panel) and in the meridional plane (Z  {\it vs} $\rm R=\sqrt{X^2+Y^2}$, left panel). Galactocentric cartesian
coordinates are designated by X, Y, and Z, Z being the height over the Galactic plane. The circle with a dot symbol ($\odot)$  in both panels shows the position of the Sun today, while the filled black star is the current position of BPM\,3066. Bottom panels: Orbital energy, E, {\it vs} the vertical component of the angular momentum, $\rm L_Z$ (left panel) and the action diamond (right panel, see text for details). BPM\,3066 is marked with the black filled stars. The background population is the `good-parallax sample' of \citet[][]{bonifacio21}, with stars classified as thin, thick, and halo stars shown as red, green, and grey dots. Gaia-Sausage-Enceladus  and Sequoia/Thamnos candidates are shown in magenta and cyan, respectively. The shaded cyan region in the action diamond shows the criteria adopted from \citet[][]{feuillet2021MNRAS.508.1489F} to select Sequoia candidates.}
    \label{fig:q53kin}
\end{figure}

The orbital history of BPM\,3066 is shown in Fig,\,\ref{fig:q53kin}. We back-integrated for 1\,Gyr the stellar orbit using the galpy code \citep[][]{bovy15}, together with Gaia mission data release 3 (DR3) astrometric data \citep[][]{gaiadr3} and radial velocity. The Gaia DR3 parallax was zero-point corrected following the prescription of \citet[][]{lindegren21}. We adopted the standard Galactic potential MWPontential2014, the \citet[][]{schoenrich10} solar peculiar velocities, a distance from the Sun of 8\,kpc, and a circular velocity at the solar distance of 220 \kms \citep[][]{bovy12}. Uncertainties were evaluated by repeating the calculations for a 1000 realisations of the input parameters, with an extraction process similar to that described in \citet[][]{bonifacio21,bonifacio24} using the pyia code \citep[][]{pyia}. 

BPM\,3066 lies at 0.45\,kpc from the Sun and 0.28\,kpc below the Galactic plane. It has an eccentric orbit with e=0.59  (see upper-right panel of Fig.\,\ref{fig:q53kin}), and reaches a maximum height over the galactic plane of 0.75\,kpc (upper-left panel) and minimum and maximum Galactocentric distances of $\rm r_{peri}=2.76$\,kpc and $\rm r_{apo}=10.62$\,kpc, respectively. Thus, its orbit is  confined to the Galactic plane, and it wanders in regions relatively far from the Galactic centre and the Bulge. The orbit of BPM\,3066 is also highly retrograde and of relatively low energy, hence it is tightly bound to the Galaxy (bottom-left panel).  

The bottom panels of Fig.\,\ref{fig:q53kin} present two of the kinematic planes most widely used to characterise stellar orbits: (i) the orbital energy, E, versus the vertical component of the angular momentum, $\rm L_Z$ (bottom-left panel), and (ii) the so-called `action-diamond', namely the difference of the vertical ($\rm J_Z$) and radial ($\rm J_R$) actions versus{\it } the azimuthal action ($\rm J_\phi=L_Z$), all normalised to the `total action' ($\rm J_{tot}=|J_\phi|+J_R+J_Z$). In both planes, besides BPM\,3066 (black star), we plotted for reference the `good-parallax sample' of \citet[][]{bonifacio21}. The stars of this sample were classified as thin disc (red), thick disc (green), and halo (grey), following the \citet[][]{bensby14} criteria. Additionally, candidate stars that belong to GSE are marked in magenta and were selected following the \citet[][]{feuillet2021MNRAS.508.1489F} criteria in the $\rm\sqrt{J_R}$ versus $\rm L_Z$ plane (which is not shown in the figure). Finally, cyan dots are candidate stars that belong to the Sequoia accretion event. The cyan rectangle shown in the action diamond (bottom-right panel) shows the criteria adopted for this selection \citep[][]{feuillet2021MNRAS.508.1489F}. Clearly, according to these criteria, BPM\,3066 may be a former member of the Sequoia galaxy. 

As pointed out by \citet[][see their Appendix C]{feuillet2021MNRAS.508.1489F}, the criteria adopted to select the Sequoia population include stars that span a wide range of energies (cyan point in the bottom-left panel), while previous selections \citep[][]{myeong19,naidu20} attributed Sequoia stars to the high-energy part only, the low-energy part being attributed to another structure named Thamnos \citep[][]{koppelman19,naidu20}. If we followed these latter works, BPM\,3066 would be classified as belonging to Thamnos. Therefore, caution must be exerted when comparing our results with the literature and the nomenclature adopted.

\citet[][]{simpson2021MNRAS.507...43S} separate different populations (GSE, retrograde and prograde halo, and disc stars) in the $\sqrt{J_R}$ versus $\rm J_\phi$ plane. The stars we select here as Sequoia candidates using the action diamond, occupy the region of the `retrograde halo' in \citet[][see their Figure 4a and 6b]{simpson2021MNRAS.507...43S}. 
Indeed, these authors identified BPM\,3066 as a lithium-rich dwarf belonging to the `retrograde halo'.

\section{Discussion}\label{sec:dis}

 From the Galactic $^9$Be-Fe relation \citep{molaro2020MNRAS.496.2902M},  
 the expected Be value for BPM\,3066 is  $\rm A(^9Be) \approx  -0.1$; 
 the   observed overabundance of beryllium above this value is therefore   2.2 dex (see Fig.\,\ref{fig:fig3}, upper panel).
A  beryllium abundance  excess as large as this
one above the Galactic trend   has never been observed in any other known star. 
Moreover, this is accompanied by a considerable excess of $^7$Li, since a
 Li abundance of A(Li) =3.0 is too high considering the stellar metallicity. The star is expected to share the $^7$Li of the Spite plateau of  about $\rm A(^7Li) = 2.2$ and therefore there is a $^7$Li overabundance of $\approx 0.8$\,dex (see Fig.\,\ref{fig:fig3}, bottom panel).
 Both overabundances   require a  mechanism to produce and accumulate Li and Be   in the stellar atmosphere.  Since  no significant internal mixing is expected in an unevolved star like BPM\,3066, it is very likely that the process is the same for both elements.  The simultaneous presence of an excess in beryllium rules out contamination from a nova or from an AGB star, which  could only  make lithium.

\subsection{Other similar stars}
A few  other notable  exceptions of Li-rich dwarf stars have been reported in the literature.
One  is  BD+23 3912, with A($^7$Li) = 2.60  and $\rm [Fe/H] =-1.45$ \citep{bonifacio1997}. However, in this star beryllium  was  found,  A($^9$Be) = --0.16, and it therefore lies on  the linear Be-Fe relation \citep{Boesgaard2011}.
Several  unevolved stars  with  an excess of $^7$Li   have been reported  by \citet{li2018ApJ...852L..31L,li22}.
One star in the globular cluster M4, N. 37934, shows A(Li)=2.87. The star is also Na rich  and \citet{monaco2011A&A...529A..90M} suggested that lithium is produced in parallel to sodium.   Unfortunately, beryllium investigation is  out of reach for this star.
A very  Li-rich dwarf  with  A($^7$Li) = 4.03  has been found in the globular cluster NGC 6397
by \citet{koch2011ApJ...738L..29K},  and  \citet{pasquini2014A&A...563A...3P}  searched unsuccessfully for the presence of beryllium.

\citet{tan2009} determined beryllium abundances for 25 metal-poor stars and one of them, star HD\,106038 (filled square in Fig.\,\ref{fig:fig3}), is very close to the case studied in this paper.
HD\,106038 shows A($^9$Be) = 1.37 $\pm$ 0.12  and A($^7$Li) = 2.55 at a metallicity of $\rm [Fe/H] = -1.3$.  \citet{Asplund2006} derived A($^7$Li) = 2.49 ([Fe/H]=-1.35) and \citet{Smiljanic2009} derived  A($^9$Be) = 1.40 for this star, in good agreement with  \citet{tan2009}. The star has a $^7$Li abundance  about 0.3 dex higher than the Spite plateau and a $^9$Be overabundance of about 1.2 dex with respect to the other stars with a similar metallicity (see Fig.\,\ref{fig:fig3}). The Galactic orbit of HD 106038  has been found to be typical for a halo star \citep{caffau2005A&A...441..533C}.

Other notable stars in the \citet{tan2009} sample are HD\,126681 with $\rm [Fe/H]  = -1.16$ and  A($^9$Be) = 0.68; HD\,132475 with [Fe/H] = --1.49 and  A($^9$Be) = 0.73; and HD 111980 with  [Fe/H] = --1.1 and A($^9$Be) = 0.52. These stars deviate moderately from the scatter of the Be-Fe relations. HD\,111980 shows A($^7$Li) =2.19  \citep{ryan1995ApJ...453..819R} but HD\,132475  has a A($^7$Li) = 2.4 from GALAH   that possibly shows a moderate enhancement.
HD 94028 with  A($^9$Be) = 0.68 as derived by  \citet{Boesgaard2006} is a potential Be rich star but it   was  remeasured  by \citet{Smiljanic2009}  who  derived a smaller Be abundance.
Following the new determination, the star is not Be enriched.

\subsection{Hypernovae as a source of $^7$Li and $^9$Be enrichment}

It has been suggested that hypernovae (HNe) 
produce large amounts of $^9$Be and $^7$Li by spallation \citep{fields2002ApJ...581..389F,nakamura2004ApJ...610..888N}.
Hypernovae are core-collapse supernovae (SNe) characterised by large kinetic energy. The energy released in the explosion can be one order of magnitude larger than that of normal core-collapse SNe \citep{iwamoto}.

\citet{fields2002ApJ...581..389F} and \citet{nakamura2004ApJ...610..888N} calculated the yields of spallation products resulting from HNe explosions.
\citet{fields2002ApJ...581..389F}  predicted a $^7$Li/$^9$Be ratio of $\approx$8.6 and \citet{nakamura2004ApJ...610..888N} predicted one of $\approx$4.2. In BPM\,3066, the ratio between the observed $^7$Li/$^9$Be is 7.9.
  In the most promising case, the yields of model (b) for SN\,1998bw by  \citet{fields2002ApJ...581..389F} ($0.451\times10^{-4} M_\odot$ and $0.671\times10^{-5} M_\odot$ for $^7$Li and $^9$Be, respectively) require a dilution with a  mass of $8\times 10^3 M_{\odot}$ of interstellar medium to produce the observed overabundances of Li and Be.  The typical mass swapped by the explosion of a core-collapse SN  is $\approx$ 10$^4 M_{\odot}$, but  the mass swapped by an HN should be larger by about a factor of 20,
 the energy being 22 times higher, according to the results of \citet{Thornton98}.  
On the other hand, the oxygen and iron yields of the HN model \citep{Woosley99} adopted by \citet{fields2002ApJ...581..389F}, respectively 2.9 and 0.5 M$_{\odot}$ if diluted in $8\times 10^3 M_{\odot}$ of pristine interstellar medium, produce A(Fe)=6.2 and A(O)=7.5, which is compatible with what is measured in our star (A(Fe)=6.0, A(O)=7.49).
The yields obtained by \citet[][$3.31\times10^{-7} M_\odot$ and $0.999\times10^{-7} M_\odot$ for $^7$Li and $^9$Be, respectively]{nakamura2004ApJ...610..888N} for SN\,1998bw are a factor of $\sim$136 (for $^7$Li) and $\sim$67 (for $^9$Be) lower than \citet{fields2002ApJ...581..389F}, and the enhancement observed in BPM 3066 cannot be reached by any reasonable dilution.

Regarding the other elements, nucleosynthetic calculations  find the ejecta of HNe to have large amounts of Si, S, and Ar \citep{nomoto2013ARA&A..51..457N}. Oxygen burning takes place in extended regions  and   more O, C, and Al are burned to produce large amounts of Si, S, and Ar. Therefore, hypernova nucleosynthesis is characterised by large abundance ratios of [(Si, S)/O]. Moreover,  higher energy explosions tend to produce larger [(Zn, Co, V)/Fe] and smaller [(Mn, Cr)/Fe] ratios compared to normal core-collapse SNe. Elements produced by $\alpha$-rich freeze-out, such as Ca, Ti, and Zn, are also enhanced.

\begin{figure}
 \includegraphics[width=0.45\textwidth]{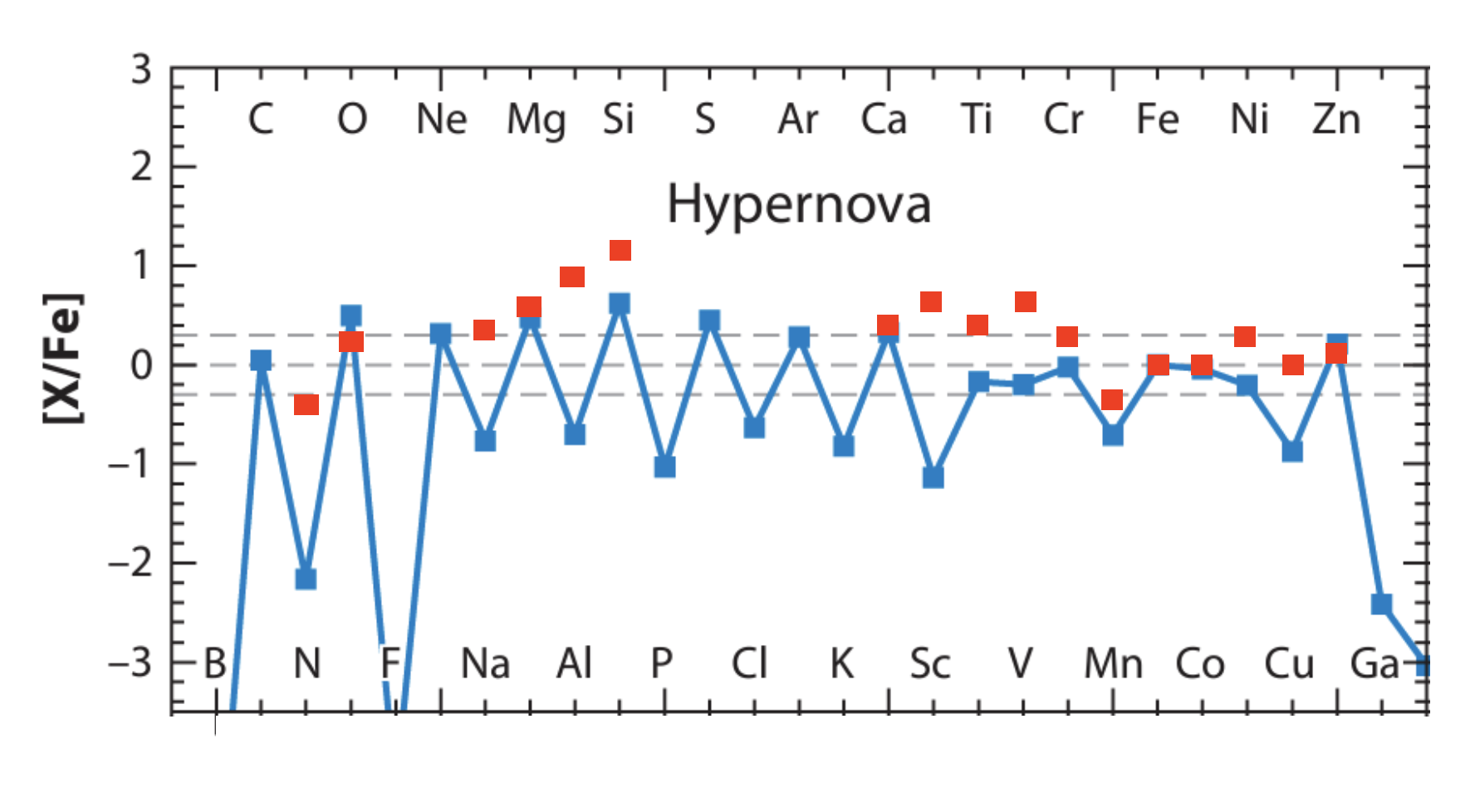}
    \caption{Observed abundances of BPM\,3066 in red filled squares compared with the hypernova yields from \citet{nomoto2013ARA&A..51..457N}.   }
    \label{fig:fig4}
\end{figure}

In Fig. \ref{fig:fig4}, the abundance pattern of BPM\,3066 is compared with the HN yields of \citet{nomoto2013ARA&A..51..457N}. 
Some stars have been reported  as showing the products of an HN \citep{skuladottir2021ApJ...915L..30S,mardini2022MNRAS.517.3993M}.
Some of the chemical predictions are indeed observed in BPM\,3066.
 An overabundance of [Si/O] is a common feature of the HNe models described in the review by \citet{nomoto2013ARA&A..51..457N}, together with moderate enhancement of Ca and Zn. In particular, the extremely high ratio of [Si/O]  ([Si/O]=1.1 using the Si abundance measured from Si\,II lines) observed in BPM\,3066 is compatible with the HN model of 25M$_{\odot}$ described in \citet{Umeda02}. 
However, the results obtained by \citet{Woosley99}, based on the HN model also used by \citet{fields2002ApJ...581..389F}, 
do not show these extreme characteristics for the ratio of Si/O. 
Moreover, 
strong deviations of the observed abundances from the model predictions also concern 
 N, Na, Al, Sc, V, and Cu. These elements are all observed to be enhanced in BPM\,3066, 
 while the HNe model of \citet{Woosley99} predicts them to be deficient. 
An HN has been suggested as being responsible for the peculiar chemical pattern of HD\,106038, which, 
along with Li and Be enhancements, shows also enhanced abundances of Si, Ni, Y, and Ba  \citep{nissen1997A&A...326..751N,smiljanic2008MNRAS.385L..93S}.
However, we emphasise that, in a similar way to BPM\,3066, in the case of HD\,106038 there are observed overabundances of Na, Sc and Ti that do not agree with the HN model predictions.

 In BPM\,3066 all the abundances of the neutron capture elements, normalised to iron, are higher than the solar ratio. 
Massive stars can produce neutron capture elements up to barium \citep{Limongi2018,Frischknecht16}, but the pattern is not compatible with the one observed in BPM\,3066, and the yields expected by massive stars are also not sufficient to explain the extreme abundances observed in this star.

To summarise, the spallation process between the CNO nuclei ejected by an 
HN and the surrounding atoms in the ISM can enhance at the expected ratio 
the abundances of lithium and beryllium in the volume swapped by the explosion. 
However, the dilution needed to reproduce the Be and Li abundances   in BPM\,3066  is not  compatible with the expected gas swapped by an HN.
On the other hand, with the same dilution and considering the HN yields, we can recover the oxygen and iron abundances measured. The chemical signatures are also puzzling. The [Si/O] is compatible with models of HNe, but strong deviations are observed for N, Na, Al, Sc, V, and Cu;
the barium and the other neutron capture elements  could be produced by a slow-process (s-process) in massive stars, but the  observed enhancement is hardly recovered.

\subsection{Planet engulfment as a source of $^7$Li and $^9$Be enrichment}

The engulfment of rocky planets with a high $^7$Li abundance has been suggested as producing $^7$Li excess \citep{siess1999MNRAS.308.1133S}.
\citet{alexander1967Obs....87..238A} proposed that the engulfing of a planet or a brown dwarf could be responsible for the Li-rich giants and this idea was then revised by  \citet{brown1989ApJS...71..293B,gratton1989A&A...215...66G,siess1999MNRAS.304..925S,siess1999MNRAS.308.1133S}, and \citet{aguilera2016ApJ...829..127A}.  BPM\,3066 is a dwarf and so the possibility that the star has produced such a Li enhancement in the course of its evolution is ruled out.

\citet{ashwell2005MNRAS.363L..81A} argued that the dwarf star J37 in the open cluster NGC\,6633,  with  A($^9$Be) = 3.0 $\pm$ 0.5 and A(Li) = 4.40 $\pm 0.14,$ might be explained by the accretion of rocky material similar to that of chondritic meteorites. Planets have now been detected around stars \citep{mayor1995Natur.378..355M} and the presence of a Li feature has been reported in the atmosphere of the two ultra hot Jupiters, WASP76b and WASP121b, together with a plethora of other ionic species \citep{tabernero2021A&A...646A.158T,borsa2021A&A...645A..24B}. The presence of Li in WASP76b has  been confirmed by the independent analyses of \citet{kesseli2022AJ....163..107K} and \citet[][]{azevedo2022A&A...666L..10A}. 

 In this context, it is interesting to note that barium was also found in WASP76b and WASP121b (strontium was also found in the latter) by using the transit technique and the extremely sensitive ESPRESSO (Echelle Spectrograph for Rocky Exoplanet and Stable Spectroscopic Observations) instrument on the ESO’s Very Large Telescope \citep{azevedo2022A&A...666L..10A}.

Lithium and beryllium are also detected in planetesimals that have been accreted onto some  white dwarfs \citep{kaiser2021Sci...371..168K,klein2021ApJ...914...61K}. 
\citet{hollands2020NatAs...4..663H} suggest  that these detections of metallic
lines in the atmospheres of white dwarfs can be explained if they have accreted fragments of planetary crusts.
\citet{doyle2021ApJ...907L..35D} argue that the high abundance of \beix cannot be accounted for either by diffusion in the outer envelope of the white dwarf or by chemical fractionation during typical rock-forming planetary formation. They suggest that \beix is the result of the spallation of oxygen atoms in icy planets by the radiation from an active star in the first million years of planetary formation.  
Thus, observations suggest that spallation processes could be the main route for both Li and Be synthesis in a planet that is later engulfed by the star. This is clearly a particular planetary evolution and different from that of the Solar System where such over abundances of beryllium are not observed.

The  likely scenario for all planets in close orbits is that they will be engulfed in the envelope of the hosting stars as the latter evolve away from the main sequence. The evolutionary routes for star-planet systems are quite complex and have been discussed in many papers \citep{livio1984MNRAS.208..763L,soker1998AJ....116.1308S}. 
For planetary masses below
a critical mass, there is a complete evaporation of the planet in the envelope.
\citet{aguilera2016ApJ...829..127A} calculate the point of dissipation in red giant stars, and find that objects with masses higher than $>$\,15 M$_J$ dissolve in the radiative interior and cannot produce an observable signature in the giant star. 
Thus, objects with masses $<$\,15 M$_J$ are the most efficient sources of Li. \citet{aguilera2016ApJ...829..127A} find that objects with a small mass can increase the superficial Li abundance of the star up to A(Li) = 2.2. 
We may expect a total amount of about $2\times10^{24}$ grams of lithium in a Jupiter-mass planet ($1.9\times10^{30}$ grams), considering a fraction of Li/H in mass of $10^{-6}$.
A complete mixing with a solar-mass star ($2\times10^{33}$ grams) would give a Li/H of $\sim10^{-9}$ in mass and $1.4\times10^{-10}$ by number or A(Li) = 2.15. 
A partial mixing would increase this number while a partial presence in the exoplanets would decrease the final Li abundance. A precise computation cannot be done but the numbers suggest that the engulfment of a planet could indeed explain the Li and Be observed.  

\citet{Smiljanic2009} rejected the planetary engulfment hypothesis for HD\,106038 arguing that there are no planets formed around metal-poor stars.
Metallicity is probably an important factor in the formation and structure of planetary systems and small planets are favoured at low metallicities \citep{buchhave2014Natur.509..593B,nielsen2023A&A...678A..74N}.   Metal-rich stars are found to be more likely  to harbour gaseous-giant planets than metal-poor stars \citep{mortie2012A&A...543A..45M,reffert2015,chavero2019MNRAS.487.3162C}. 
However, the most metal-poor stars that host a confirmed
planet, according to the Encyclopaedia of Exoplanetary Systems\footnote{\url{https://exoplanet.eu}}
, have metallicities of about --2.0 \citep[V 894 Cyg,][]{liqian} and --1.4  \citep[G 178-27,][]{latham1992}.
The number of candidate planets  with a metallicity below --1.0 is much larger, 
so that it seems safe to conclude that planets around stars with a metallicity less than --1.0
exist, so the argument of \citet{Smiljanic2009} can be rejected on the basis
of our current understanding of the properties of the stars that host an exoplanet.
However, the lowest metallicity at which planets can form remains open from a theoretical perspective. Simulations by \citet{Eemsenhuber2023EPJP..138..181E} indicate that super-Earth and sub-Neptunes should be more frequent than what has been observed at low metallicities.
The radii of discs that surround stars are predicted to decrease with decreasing metallicity, and planets are therefore expected to orbit closer to their parent stars, which favours their engulfment \citep{elsender2021MNRAS.508.5279E}. 
We note that \citet{santos2002A&A...386.1028S} studied beryllium abundances in stars both with and without planets without finding any differences. This implies that engulfing of planets that are rich in light elements is rare, at least among
high metallicity stars, as in the sample of \citet{santos2002A&A...386.1028S}.

\section{Conclusions}\label{sec:sum}

We have identified a star, BPM\,3066, with extremely high Li and Be abundances.
The observed ratio between the two elements strongly suggests a common origin through spallation processes. The elemental synthesis could have occurred in the environment of an hypernova. This is supported by some additional abundance anomalies besides Li and Be, such as the high value of Si observed in the star, but other elements are found to be abundant while they should be deficient according to HNe predictions.

Alternatively, the high Li and Be abundances could result from the engulfing of rocky planets where the light elements have been formed by spallation processes in the icy surfaces of the planets under the action of energetic radiation of the forming star.
In the hypothesis that BPM\,3066 is the result of  a planet engulfment, this would provide further 
evidence that  planets can form  at metallicities of [Fe/H]$<-1$, adding to the existing evidence from already detected planets around low-metallicity stars
and the large number of candidate planets. In any case, it is  intriguing that such an extreme star   has been found  to belong to the Sequoia or Thamnos accreted galaxy.

\section*{Data availability}
The lines used for the chemical abundance analysis are only available in electronic form at the CDS via anonymous ftp to cdsarc.u-strasbg.fr (130.79.128.5) or via http://cdsarc.cds.unistra.fr/viz-bin/cat/J/A+A/vol/page.

\begin{acknowledgements}
We thank the anonymous referee for a constructive report, which helped improve the presentation.
NOIRLab IRAF is distributed by the Community Science and Data Center at NSF NOIRLab, which is managed by the Association of Universities for Research in Astronomy (AURA) under a cooperative agreement with the U.S. National Science Foundation.  
This research has made use of data obtained from or tools provided by the portal exoplanet.eu of The Extrasolar
Planets Encyclopaedia.
This publication makes use of VOSA, developed under the Spanish Virtual Observatory (https://svo.cab.inta-csic.es) project funded by MCIN/AEI/10.13039/501100011033/ through grant PID2020-112949GB-I00.
VOSA has been partially updated by using funding from the European Union's Horizon 2020 Research and Innovation Programme, under Grant Agreement nº 776403 (EXOPLANETS-A).
PB acknowledges support   from the ERC advanced grant N. 835087 -- SPIAKID. 
GC acknowledges the grant PRIN
project No. 2022X4TM3H `Cosmic POT' from Ministero dell'Universit\`a e della
Ricerca (MUR).  GC also thanks for the support  INAF  (Large Grant 2023, 
EPOCH) and the European Union (ChETEC-INFRA, project no. 101008324).
   
\end{acknowledgements}

\bibliography{sequoia_v3}

\begin{thebibliography}{119}
\expandafter\ifx\csname natexlab\endcsname\relax\def\natexlab#1{#1}\fi

\bibitem[{{Aguilera-G{\'o}mez} {et~al.}(2016){Aguilera-G{\'o}mez},
  {Chanam{\'e}}, {Pinsonneault}, \& {Carlberg}}]{aguilera2016ApJ...829..127A}
{Aguilera-G{\'o}mez}, C., {Chanam{\'e}}, J., {Pinsonneault}, M.~H., \&
  {Carlberg}, J.~K. 2016, \apj, 829, 127

\bibitem[{{Alexander}(1967)}]{alexander1967Obs....87..238A}
{Alexander}, J.~B. 1967, The Observatory, 87, 238

\bibitem[{{Alvarez} \& {Plez}(1998)}]{alvarezplez}
{Alvarez}, R. \& {Plez}, B. 1998, \aap, 330, 1109

\bibitem[{{Amarsi} {et~al.}(2016){Amarsi}, {Lind}, {Asplund}, {Barklem}, \&
  {Collet}}]{amarsi2016}
{Amarsi}, A.~M., {Lind}, K., {Asplund}, M., {Barklem}, P.~S., \& {Collet}, R.
  2016, \mnras, 463, 1518

\bibitem[{{Ashwell} {et~al.}(2005){Ashwell}, {Jeffries}, {Smalley},
  {Deliyannis}, {Steinhauer}, \& {King}}]{ashwell2005MNRAS.363L..81A}
{Ashwell}, J.~F., {Jeffries}, R.~D., {Smalley}, B., {et~al.} 2005, \mnras, 363,
  L81

\bibitem[{{Asplund} {et~al.}(2006){Asplund}, {Lambert}, {Nissen}, {Primas}, \&
  {Smith}}]{Asplund2006}
{Asplund}, M., {Lambert}, D.~L., {Nissen}, P.~E., {Primas}, F., \& {Smith},
  V.~V. 2006, \apj, 644, 229

\bibitem[{{Azevedo Silva} {et~al.}(2022){Azevedo Silva}, {Demangeon}, {Santos},
  {Allart}, {Borsa}, {Cristo}, {Esparza-Borges}, {Seidel}, {Palle}, {Sousa},
  {Tabernero}, {Zapatero Osorio}, {Cristiani}, {Pepe}, {Rebolo}, {Adibekyan},
  {Alibert}, {Barros}, {Bouchy}, {Bourrier}, {Lo Curto}, {Di Marcantonio},
  {D'Odorico}, {Ehrenreich}, {Figueira}, {Gonz{\'a}lez Hern{\'a}ndez}, {Lovis},
  {Martins}, {Mehner}, {Micela}, {Molaro}, {Mounzer}, {Nunes}, {Sozzetti},
  {Su{\'a}rez Mascare{\~n}o}, \& {Udry}}]{azevedo2022A&A...666L..10A}
{Azevedo Silva}, T., {Demangeon}, O.~D.~S., {Santos}, N.~C., {et~al.} 2022,
  \aap, 666, L10

\bibitem[{{Barb{\'a}} {et~al.}(2019){Barb{\'a}}, {Minniti}, {Geisler},
  {Alonso-Garc{\'\i}a}, {Hempel}, {Monachesi}, {Arias}, \&
  {G{\'o}mez}}]{barba19}
{Barb{\'a}}, R.~H., {Minniti}, D., {Geisler}, D., {et~al.} 2019, \apjl, 870,
  L24

\bibitem[{{Barklem}(2016)}]{barklem2016}
{Barklem}, P.~S. 2016, \pra, 93, 042705

\bibitem[{{Bayo} {et~al.}(2008){Bayo}, {Rodrigo}, {Barrado Y Navascu{\'e}s},
  {Solano}, {Guti{\'e}rrez}, {Morales-Calder{\'o}n}, \& {Allard}}]{vosa08}
{Bayo}, A., {Rodrigo}, C., {Barrado Y Navascu{\'e}s}, D., {et~al.} 2008, \aap,
  492, 277

\bibitem[{{Belokurov} {et~al.}(2018){Belokurov}, {Erkal}, {Evans}, {Koposov},
  \& {Deason}}]{Belokurov2018}
{Belokurov}, V., {Erkal}, D., {Evans}, N.~W., {Koposov}, S.~E., \& {Deason},
  A.~J. 2018, \mnras, 478, 611

\bibitem[{{Bensby} {et~al.}(2014){Bensby}, {Feltzing}, \& {Oey}}]{bensby14}
{Bensby}, T., {Feltzing}, S., \& {Oey}, M.~S. 2014, \aap, 562, A71

\bibitem[{{Bergemann}(2011)}]{bergemann2011}
{Bergemann}, M. 2011, \mnras, 413, 2184

\bibitem[{{Bergemann} \& {Cescutti}(2010)}]{bergemann2010}
{Bergemann}, M. \& {Cescutti}, G. 2010, \aap, 522, A9

\bibitem[{{Bergemann} {et~al.}(2017){Bergemann}, {Collet}, {Amarsi}, {Kovalev},
  {Ruchti}, \& {Magic}}]{bergemann2017}
{Bergemann}, M., {Collet}, R., {Amarsi}, A.~M., {et~al.} 2017, \apj, 847, 15

\bibitem[{{Bergemann} {et~al.}(2013){Bergemann}, {Kudritzki}, {W{\"u}rl},
  {Plez}, {Davies}, \& {Gazak}}]{bergemann2013}
{Bergemann}, M., {Kudritzki}, R.-P., {W{\"u}rl}, M., {et~al.} 2013, \apj, 764,
  115

\bibitem[{{Bergemann} {et~al.}(2012){Bergemann}, {Lind}, {Collet}, {Magic}, \&
  {Asplund}}]{bergemann2012}
{Bergemann}, M., {Lind}, K., {Collet}, R., {Magic}, Z., \& {Asplund}, M. 2012,
  \mnras, 427, 27

\bibitem[{{Bergemann} {et~al.}(2010){Bergemann}, {Pickering}, \&
  {Gehren}}]{bergemannco2010}
{Bergemann}, M., {Pickering}, J.~C., \& {Gehren}, T. 2010, \mnras, 401, 1334

\bibitem[{{Boesgaard} \& {Novicki}(2006)}]{Boesgaard2006}
{Boesgaard}, A.~M. \& {Novicki}, M.~C. 2006, \apj, 641, 1122

\bibitem[{{Boesgaard} {et~al.}(2011){Boesgaard}, {Rich}, {Levesque}, \&
  {Bowler}}]{Boesgaard2011}
{Boesgaard}, A.~M., {Rich}, J.~A., {Levesque}, E.~M., \& {Bowler}, B.~P. 2011,
  \apj, 743, 140

\bibitem[{{Bonifacio} {et~al.}(2024){Bonifacio}, {Caffau}, {Monaco},
  {Sbordone}, {Spite}, {Mucciarelli}, {Fran{\c{c}}ois}, {Lombardo}, \& {Matas
  Pinto}}]{bonifacio24}
{Bonifacio}, P., {Caffau}, E., {Monaco}, L., {et~al.} 2024, \aap, 684, A91

\bibitem[{{Bonifacio} \& {Molaro}(1997)}]{bonifacio1997}
{Bonifacio}, P. \& {Molaro}, P. 1997, \mnras, 285, 847

\bibitem[{{Bonifacio} {et~al.}(2021){Bonifacio}, {Monaco}, {Salvadori},
  {Caffau}, {Spite}, {Sbordone}, {Spite}, {Ludwig}, {Di Matteo}, {Haywood},
  {Fran{\c{c}}ois}, {Koch-Hansen}, {Christlieb}, \& {Zaggia}}]{bonifacio21}
{Bonifacio}, P., {Monaco}, L., {Salvadori}, S., {et~al.} 2021, \aap, 651, A79

\bibitem[{{Borsa} {et~al.}(2021){Borsa}, {Allart}, {Casasayas-Barris},
  {Tabernero}, {Zapatero Osorio}, {Cristiani}, {Pepe}, {Rebolo}, {Santos},
  {Adibekyan}, {Bourrier}, {Demangeon}, {Ehrenreich}, {Pall{\'e}}, {Sousa},
  {Lillo-Box}, {Lovis}, {Micela}, {Oshagh}, {Poretti}, {Sozzetti}, {Allende
  Prieto}, {Alibert}, {Amate}, {Benz}, {Bouchy}, {Cabral}, {Dekker},
  {D'Odorico}, {Di Marcantonio}, {Figueira}, {Genova Santos}, {Gonz{\'a}lez
  Hern{\'a}ndez}, {Lo Curto}, {Manescau}, {Martins}, {M{\'e}gevand}, {Mehner},
  {Molaro}, {Nunes}, {Riva}, {Su{\'a}rez Mascare{\~n}o}, {Udry}, \&
  {Zerbi}}]{borsa2021A&A...645A..24B}
{Borsa}, F., {Allart}, R., {Casasayas-Barris}, N., {et~al.} 2021, \aap, 645,
  A24

\bibitem[{{Bovy}(2015)}]{bovy15}
{Bovy}, J. 2015, \apjs, 216, 29

\bibitem[{{Bovy} {et~al.}(2012){Bovy}, {Allende Prieto}, {Beers}, {Bizyaev},
  {da Costa}, {Cunha}, {Ebelke}, {Eisenstein}, {Frinchaboy}, {Garc{\'\i}a
  P{\'e}rez}, {Girardi}, {Hearty}, {Hogg}, {Holtzman}, {Maia}, {Majewski},
  {Malanushenko}, {Malanushenko}, {M{\'e}sz{\'a}ros}, {Nidever}, {O'Connell},
  {O'Donnell}, {Oravetz}, {Pan}, {Rocha-Pinto}, {Schiavon}, {Schneider},
  {Schultheis}, {Skrutskie}, {Smith}, {Weinberg}, {Wilson}, \&
  {Zasowski}}]{bovy12}
{Bovy}, J., {Allende Prieto}, C., {Beers}, T.~C., {et~al.} 2012, \apj, 759, 131

\bibitem[{{Bressan} {et~al.}(2012){Bressan}, {Marigo}, {Girardi}, {Salasnich},
  {Dal Cero}, {Rubele}, \& {Nanni}}]{bressan12}
{Bressan}, A., {Marigo}, P., {Girardi}, L., {et~al.} 2012, \mnras, 427, 127

\bibitem[{{Brown} {et~al.}(1989){Brown}, {Sneden}, {Lambert}, \&
  {Dutchover}}]{brown1989ApJS...71..293B}
{Brown}, J.~A., {Sneden}, C., {Lambert}, D.~L., \& {Dutchover}, Edward, J.
  1989, \apjs, 71, 293

\bibitem[{{Buchhave} {et~al.}(2014){Buchhave}, {Bizzarro}, {Latham},
  {Sasselov}, {Cochran}, {Endl}, {Isaacson}, {Juncher}, \&
  {Marcy}}]{buchhave2014Natur.509..593B}
{Buchhave}, L.~A., {Bizzarro}, M., {Latham}, D.~W., {et~al.} 2014, \nat, 509,
  593

\bibitem[{{Buder} {et~al.}(2021){Buder}, {Sharma}, {Kos}, {Amarsi},
  {Nordlander}, {Lind}, {Martell}, {Asplund}, {Bland-Hawthorn}, {Casey}, {de
  Silva}, {D'Orazi}, {Freeman}, {Hayden}, {Lewis}, {Lin}, {Schlesinger},
  {Simpson}, {Stello}, {Zucker}, {Zwitter}, {Beeson}, {Buck}, {Casagrande},
  {Clark}, {{\v{C}}otar}, {da Costa}, {de Grijs}, {Feuillet}, {Horner},
  {Kafle}, {Khanna}, {Kobayashi}, {Liu}, {Montet}, {Nandakumar}, {Nataf},
  {Ness}, {Spina}, {Tepper-Garc{\'\i}a}, {Ting}, {Traven},
  {Vogrin{\v{c}}i{\v{c}}}, {Wittenmyer}, {Wyse}, {{\v{Z}}erjal}, \& {Galah
  Collaboration}}]{buder2021MNRAS.506..150B}
{Buder}, S., {Sharma}, S., {Kos}, J., {et~al.} 2021, \mnras, 506, 150

\bibitem[{{Caffau} {et~al.}(2005){Caffau}, {Bonifacio}, {Faraggiana},
  {Fran{\c{c}}ois}, {Gratton}, \& {Barbieri}}]{caffau2005A&A...441..533C}
{Caffau}, E., {Bonifacio}, P., {Faraggiana}, R., {et~al.} 2005, \aap, 441, 533

\bibitem[{{Caffau} {et~al.}(2024){Caffau}, {Bonifacio}, {Monaco}, {Steffen},
  {Sbordone}, {Spite}, {Fran{\c{c}}ois}, {Gallagher}, {Ludwig}, \&
  {Molaro}}]{caffau_sdss_2024}
{Caffau}, E., {Bonifacio}, P., {Monaco}, L., {et~al.} 2024, \aap, 691, A245

\bibitem[{{Caffau} {et~al.}(2011{\natexlab{a}}){Caffau}, {Faraggiana},
  {Ludwig}, {Bonifacio}, \& {Steffen}}]{caffau11a}
{Caffau}, E., {Faraggiana}, R., {Ludwig}, H.~G., {Bonifacio}, P., \& {Steffen},
  M. 2011{\natexlab{a}}, Astronomische Nachrichten, 332, 128

\bibitem[{{Caffau} {et~al.}(2011{\natexlab{b}}){Caffau}, {Ludwig}, {Steffen},
  {Freytag}, \& {Bonifacio}}]{caffau11b}
{Caffau}, E., {Ludwig}, H.~G., {Steffen}, M., {Freytag}, B., \& {Bonifacio}, P.
  2011{\natexlab{b}}, \solphys, 268, 255

\bibitem[{{Chavero} {et~al.}(2019){Chavero}, {de la Reza}, {Ghezzi}, {Llorente
  de Andr{\'e}s}, {Pereira}, {Giuppone}, \&
  {Pinz{\'o}n}}]{chavero2019MNRAS.487.3162C}
{Chavero}, C., {de la Reza}, R., {Ghezzi}, L., {et~al.} 2019, \mnras, 487, 3162

\bibitem[{{Coc} \& {Vangioni}(2017)}]{coc17}
{Coc}, A. \& {Vangioni}, E. 2017, International Journal of Modern Physics E,
  26, 1741002

\bibitem[{{Dekker} {et~al.}(2000){Dekker}, {D'Odorico}, {Kaufer}, {Delabre}, \&
  {Kotzlowski}}]{dekker00}
{Dekker}, H., {D'Odorico}, S., {Kaufer}, A., {Delabre}, B., \& {Kotzlowski}, H.
  2000, in Society of Photo-Optical Instrumentation Engineers (SPIE) Conference
  Series, Vol. 4008, Optical and IR Telescope Instrumentation and Detectors,
  ed. M.~{Iye} \& A.~F. {Moorwood}, 534--545

\bibitem[{{Doyle} {et~al.}(2021){Doyle}, {Desch}, \&
  {Young}}]{doyle2021ApJ...907L..35D}
{Doyle}, A.~E., {Desch}, S.~J., \& {Young}, E.~D. 2021, \apjl, 907, L35

\bibitem[{{Drawin}(1968)}]{drawin1968}
{Drawin}, H.-W. 1968, Zeitschrift fur Physik, 211, 404

\bibitem[{{Drawin}(1969)}]{drawin1969}
{Drawin}, H.~W. 1969, Zeitschrift fur Physik, 225, 483

\bibitem[{{Elsender} \& {Bate}(2021)}]{elsender2021MNRAS.508.5279E}
{Elsender}, D. \& {Bate}, M.~R. 2021, \mnras, 508, 5279

\bibitem[{{Emsenhuber} {et~al.}(2023){Emsenhuber}, {Mordasini}, \&
  {Burn}}]{Eemsenhuber2023EPJP..138..181E}
{Emsenhuber}, A., {Mordasini}, C., \& {Burn}, R. 2023, European Physical
  Journal Plus, 138, 181

\bibitem[{{Feuillet} {et~al.}(2021){Feuillet}, {Sahlholdt}, {Feltzing}, \&
  {Casagrande}}]{feuillet2021MNRAS.508.1489F}
{Feuillet}, D.~K., {Sahlholdt}, C.~L., {Feltzing}, S., \& {Casagrande}, L.
  2021, \mnras, 508, 1489

\bibitem[{{Fields} {et~al.}(2002){Fields}, {Daigne}, {Cass{\'e}}, \&
  {Vangioni-Flam}}]{fields2002ApJ...581..389F}
{Fields}, B.~D., {Daigne}, F., {Cass{\'e}}, M., \& {Vangioni-Flam}, E. 2002,
  \apj, 581, 389

\bibitem[{{Fitzpatrick} {et~al.}(2024){Fitzpatrick}, {Placco}, {Bolton},
  {Merino}, {Ridgway}, \& {Stanghellini}}]{fitzpatrick24}
{Fitzpatrick}, M., {Placco}, V., {Bolton}, A., {et~al.} 2024, arXiv e-prints,
  arXiv:2401.01982

\bibitem[{{Frischknecht} {et~al.}(2016){Frischknecht}, {Hirschi}, {Pignatari},
  {Maeder}, {Meynet}, {Chiappini}, {Thielemann}, {Rauscher}, {Georgy}, \&
  {Ekstr{\"o}m}}]{Frischknecht16}
{Frischknecht}, U., {Hirschi}, R., {Pignatari}, M., {et~al.} 2016, \mnras, 456,
  1803

\bibitem[{{Gaia Collaboration} {et~al.}(2023){Gaia Collaboration}, {Vallenari},
  {Brown}, {Prusti}, {de Bruijne}, {Arenou}, {Babusiaux}, {Biermann},
  {Creevey}, {Ducourant}, {Evans}, {Eyer}, {Guerra}, {Hutton}, {Jordi},
  {Klioner}, {Lammers}, {Lindegren}, {Luri}, {Mignard}, {Panem}, {Pourbaix},
  {Randich}, {Sartoretti}, {Soubiran}, {Tanga}, {Walton}, {Bailer-Jones},
  {Bastian}, {Drimmel}, {Jansen}, {Katz}, {Lattanzi}, {van Leeuwen}, {Bakker},
  {Cacciari}, {Casta{\~n}eda}, {De Angeli}, {Fabricius}, {Fouesneau},
  {Fr{\'e}mat}, {Galluccio}, {Guerrier}, {Heiter}, {Masana}, {Messineo},
  {Mowlavi}, {Nicolas}, {Nienartowicz}, {Pailler}, {Panuzzo}, {Riclet}, {Roux},
  {Seabroke}, {Sordo}, {Th{\'e}venin}, {Gracia-Abril}, {Portell}, {Teyssier},
  {Altmann}, {Andrae}, {Audard}, {Bellas-Velidis}, {Benson}, {Berthier},
  {Blomme}, {Burgess}, {Busonero}, {Busso}, {C{\'a}novas}, {Carry}, {Cellino},
  {Cheek}, {Clementini}, {Damerdji}, {Davidson}, {de Teodoro}, {Nu{\~n}ez
  Campos}, {Delchambre}, {Dell'Oro}, {Esquej}, {Fern{\'a}ndez-Hern{\'a}ndez},
  {Fraile}, {Garabato}, {Garc{\'\i}a-Lario}, {Gosset}, {Haigron}, {Halbwachs},
  {Hambly}, {Harrison}, {Hern{\'a}ndez}, {Hestroffer}, {Hodgkin}, {Holl},
  {Jan{\ss}en}, {Jevardat de Fombelle}, {Jordan}, {Krone-Martins}, {Lanzafame},
  {L{\"o}ffler}, {Marchal}, {Marrese}, {Moitinho}, {Muinonen}, {Osborne},
  {Pancino}, {Pauwels}, {Recio-Blanco}, {Reyl{\'e}}, {Riello}, {Rimoldini},
  {Roegiers}, {Rybizki}, {Sarro}, {Siopis}, {Smith}, {Sozzetti}, {Utrilla},
  {van Leeuwen}, {Abbas}, {{\'A}brah{\'a}m}, {Abreu Aramburu}, {Aerts},
  {Aguado}, {Ajaj}, {Aldea-Montero}, {Altavilla}, {{\'A}lvarez}, {Alves},
  {Anders}, {Anderson}, {Anglada Varela}, {Antoja}, {Baines}, {Baker},
  {Balaguer-N{\'u}{\~n}ez}, {Balbinot}, {Balog}, {Barache}, {Barbato},
  {Barros}, {Barstow}, {Bartolom{\'e}}, {Bassilana}, {Bauchet}, {Becciani},
  {Bellazzini}, {Berihuete}, {Bernet}, {Bertone}, {Bianchi}, {Binnenfeld},
  {Blanco-Cuaresma}, {Blazere}, {Boch}, {Bombrun}, {Bossini}, {Bouquillon},
  {Bragaglia}, {Bramante}, {Breedt}, {Bressan}, {Brouillet}, {Brugaletta},
  {Bucciarelli}, {Burlacu}, {Butkevich}, {Buzzi}, {Caffau}, {Cancelliere},
  {Cantat-Gaudin}, {Carballo}, {Carlucci}, {Carnerero}, {Carrasco},
  {Casamiquela}, {Castellani}, {Castro-Ginard}, {Chaoul}, {Charlot}, {Chemin},
  {Chiaramida}, {Chiavassa}, {Chornay}, {Comoretto}, {Contursi}, {Cooper},
  {Cornez}, {Cowell}, {Crifo}, {Cropper}, {Crosta}, {Crowley}, {Dafonte},
  {Dapergolas}, {David}, {David}, {de Laverny}, {De Luise}, {De March}, {De
  Ridder}, {de Souza}, {de Torres}, {del Peloso}, {del Pozo}, {Delbo},
  {Delgado}, {Delisle}, {Demouchy}, {Dharmawardena}, {Di Matteo}, {Diakite},
  {Diener}, {Distefano}, {Dolding}, {Edvardsson}, {Enke}, {Fabre}, {Fabrizio},
  {Faigler}, {Fedorets}, {Fernique}, {Fienga}, {Figueras}, {Fournier},
  {Fouron}, {Fragkoudi}, {Gai}, {Garcia-Gutierrez}, {Garcia-Reinaldos},
  {Garc{\'\i}a-Torres}, {Garofalo}, {Gavel}, {Gavras}, {Gerlach}, {Geyer},
  {Giacobbe}, {Gilmore}, {Girona}, {Giuffrida}, {Gomel}, {Gomez},
  {Gonz{\'a}lez-N{\'u}{\~n}ez}, {Gonz{\'a}lez-Santamar{\'\i}a},
  {Gonz{\'a}lez-Vidal}, {Granvik}, {Guillout}, {Guiraud},
  {Guti{\'e}rrez-S{\'a}nchez}, {Guy}, {Hatzidimitriou}, {Hauser}, {Haywood},
  {Helmer}, {Helmi}, {Sarmiento}, {Hidalgo}, {Hilger}, {H{\l}adczuk}, {Hobbs},
  {Holland}, {Huckle}, {Jardine}, {Jasniewicz}, {Jean-Antoine Piccolo},
  {Jim{\'e}nez-Arranz}, {Jorissen}, {Juaristi Campillo}, {Julbe}, {Karbevska},
  {Kervella}, {Khanna}, {Kontizas}, {Kordopatis}, {Korn}, {K{\'o}sp{\'a}l},
  {Kostrzewa-Rutkowska}, {Kruszy{\'n}ska}, {Kun}, {Laizeau}, {Lambert},
  {Lanza}, {Lasne}, {Le Campion}, {Lebreton}, {Lebzelter}, {Leccia}, {Leclerc},
  {Lecoeur-Taibi}, {Liao}, {Licata}, {Lindstr{\o}m}, {Lister}, {Livanou},
  {Lobel}, {Lorca}, {Loup}, {Madrero Pardo}, {Magdaleno Romeo}, {Managau},
  {Mann}, {Manteiga}, {Marchant}, {Marconi}, {Marcos}, {Marcos Santos},
  {Mar{\'\i}n Pina}, {Marinoni}, {Marocco}, {Marshall}, {Martin Polo},
  {Mart{\'\i}n-Fleitas}, {Marton}, {Mary}, {Masip}, {Massari},
  {Mastrobuono-Battisti}, {Mazeh}, {McMillan}, {Messina}, {Michalik}, {Millar},
  {Mints}, {Molina}, {Molinaro}, {Moln{\'a}r}, {Monari}, {Mongui{\'o}},
  {Montegriffo}, {Montero}, {Mor}, {Mora}, {Morbidelli}, {Morel}, {Morris},
  {Muraveva}, {Murphy}, {Musella}, {Nagy}, {Noval}, {Oca{\~n}a}, {Ogden},
  {Ordenovic}, {Osinde}, {Pagani}, {Pagano}, {Palaversa}, {Palicio},
  {Pallas-Quintela}, {Panahi}, {Payne-Wardenaar}, {Pe{\~n}alosa Esteller},
  {Penttil{\"a}}, {Pichon}, {Piersimoni}, {Pineau}, {Plachy}, {Plum}, {Poggio},
  {Pr{\v{s}}a}, {Pulone}, {Racero}, {Ragaini}, {Rainer}, {Raiteri}, {Rambaux},
  {Ramos}, {Ramos-Lerate}, {Re Fiorentin}, {Regibo}, {Richards}, {Rios Diaz},
  {Ripepi}, {Riva}, {Rix}, {Rixon}, {Robichon}, {Robin}, {Robin}, {Roelens},
  {Rogues}, {Rohrbasser}, {Romero-G{\'o}mez}, {Rowell}, {Royer}, {Ruz Mieres},
  {Rybicki}, {Sadowski}, {S{\'a}ez N{\'u}{\~n}ez}, {Sagrist{\`a} Sell{\'e}s},
  {Sahlmann}, {Salguero}, {Samaras}, {Sanchez Gimenez}, {Sanna},
  {Santove{\~n}a}, {Sarasso}, {Schultheis}, {Sciacca}, {Segol}, {Segovia},
  {S{\'e}gransan}, {Semeux}, {Shahaf}, {Siddiqui}, {Siebert}, {Siltala},
  {Silvelo}, {Slezak}, {Slezak}, {Smart}, {Snaith}, {Solano}, {Solitro},
  {Souami}, {Souchay}, {Spagna}, {Spina}, {Spoto}, {Steele},
  {Steidelm{\"u}ller}, {Stephenson}, {S{\"u}veges}, {Surdej}, {Szabados},
  {Szegedi-Elek}, {Taris}, {Taylor}, {Teixeira}, {Tolomei}, {Tonello}, {Torra},
  {Torra}, {Torralba Elipe}, {Trabucchi}, {Tsounis}, {Turon}, {Ulla}, {Unger},
  {Vaillant}, {van Dillen}, {van Reeven}, {Vanel}, {Vecchiato}, {Viala},
  {Vicente}, {Voutsinas}, {Weiler}, {Wevers}, {Wyrzykowski}, {Yoldas}, {Yvard},
  {Zhao}, {Zorec}, {Zucker}, \& {Zwitter}}]{gaiadr3}
{Gaia Collaboration}, {Vallenari}, A., {Brown}, A.~G.~A., {et~al.} 2023, \aap,
  674, A1

\bibitem[{{Gratton} \& {D'Antona}(1989)}]{gratton1989A&A...215...66G}
{Gratton}, R.~G. \& {D'Antona}, F. 1989, \aap, 215, 66

\bibitem[{{Haywood} {et~al.}(2018){Haywood}, {Di Matteo}, {Lehnert}, {Snaith},
  {Khoperskov}, \& {G{\'o}mez}}]{Haywood2018}
{Haywood}, M., {Di Matteo}, P., {Lehnert}, M.~D., {et~al.} 2018, \apj, 863, 113

\bibitem[{{Helmi} {et~al.}(2018){Helmi}, {Babusiaux}, {Koppelman}, {Massari},
  {Veljanoski}, \& {Brown}}]{Helmi2018}
{Helmi}, A., {Babusiaux}, C., {Koppelman}, H.~H., {et~al.} 2018, \nat, 563, 85

\bibitem[{{Hollands} {et~al.}(2020){Hollands}, {Tremblay}, {G{\"a}nsicke},
  {Camisassa}, {Koester}, {Aungwerojwit}, {Chote}, {C{\'o}rsico}, {Dhillon},
  {Gentile-Fusillo}, {Hoskin}, {Izquierdo}, {Marsh}, \&
  {Steeghs}}]{hollands2020NatAs...4..663H}
{Hollands}, M.~A., {Tremblay}, P.~E., {G{\"a}nsicke}, B.~T., {et~al.} 2020,
  Nature Astronomy, 4, 663

\bibitem[{{Iwamoto} {et~al.}(2003){Iwamoto}, {Nomoto}, {Mazzali}, {Nakamura},
  \& {Maeda}}]{iwamoto}
{Iwamoto}, K., {Nomoto}, K., {Mazzali}, P.~A., {Nakamura}, T., \& {Maeda}, K.
  2003, in Supernovae and Gamma-Ray Bursters, ed. K.~{Weiler}, Vol. 598,
  243--281

\bibitem[{{Kaiser} {et~al.}(2021){Kaiser}, {Clemens}, {Blouin}, {Dufour},
  {Hegedus}, {Reding}, \& {B{\'e}dard}}]{kaiser2021Sci...371..168K}
{Kaiser}, B.~C., {Clemens}, J.~C., {Blouin}, S., {et~al.} 2021, Science, 371,
  168

\bibitem[{{Kesseli} {et~al.}(2022){Kesseli}, {Snellen}, {Casasayas-Barris},
  {Molli{\`e}re}, \& {S{\'a}nchez-L{\'o}pez}}]{kesseli2022AJ....163..107K}
{Kesseli}, A.~Y., {Snellen}, I.~A.~G., {Casasayas-Barris}, N., {Molli{\`e}re},
  P., \& {S{\'a}nchez-L{\'o}pez}, A. 2022, \aj, 163, 107

\bibitem[{{Klein} {et~al.}(2021){Klein}, {Doyle}, {Zuckerman}, {Dufour},
  {Blouin}, {Melis}, {Weinberger}, \& {Young}}]{klein2021ApJ...914...61K}
{Klein}, B.~L., {Doyle}, A.~E., {Zuckerman}, B., {et~al.} 2021, \apj, 914, 61

\bibitem[{{Koch} {et~al.}(2011){Koch}, {Lind}, \&
  {Rich}}]{koch2011ApJ...738L..29K}
{Koch}, A., {Lind}, K., \& {Rich}, R.~M. 2011, \apjl, 738, L29

\bibitem[{{Koppelman} {et~al.}(2019){Koppelman}, {Helmi}, {Massari},
  {Price-Whelan}, \& {Starkenburg}}]{koppelman19}
{Koppelman}, H.~H., {Helmi}, A., {Massari}, D., {Price-Whelan}, A.~M., \&
  {Starkenburg}, T.~K. 2019, \aap, 631, L9

\bibitem[{{Kurucz}(2005)}]{Kurucz2005}
{Kurucz}, R.~L. 2005, Memorie della Societa Astronomica Italiana Supplementi,
  8, 14

\bibitem[{{Latham} {et~al.}(1992){Latham}, {Mazeh}, {Stefanik}, {Davis},
  {Carney}, {Krymolowski}, {Laird}, {Torres}, \& {Morse}}]{latham1992}
{Latham}, D.~W., {Mazeh}, T., {Stefanik}, R.~P., {et~al.} 1992, \aj, 104, 774

\bibitem[{{Li} {et~al.}(2018){Li}, {Aoki}, {Matsuno}, {Bharat Kumar}, {Shi},
  {Suda}, \& {Zhao}}]{li2018ApJ...852L..31L}
{Li}, H., {Aoki}, W., {Matsuno}, T., {et~al.} 2018, \apjl, 852, L31

\bibitem[{{Li} {et~al.}(2022){Li}, {Aoki}, {Matsuno}, {Xing}, {Suda},
  {Tominaga}, {Chen}, {Honda}, {Ishigaki}, {Shi}, {Zhao}, \& {Zhao}}]{li22}
{Li}, H., {Aoki}, W., {Matsuno}, T., {et~al.} 2022, \apj, 931, 147

\bibitem[{{Li} \& {Qian}(2014)}]{liqian}
{Li}, L.~J. \& {Qian}, S.~B. 2014, \mnras, 444, 600

\bibitem[{{Limongi} \& {Chieffi}(2018)}]{Limongi2018}
{Limongi}, M. \& {Chieffi}, A. 2018, \apjs, 237, 13

\bibitem[{{Lindegren} {et~al.}(2021){Lindegren}, {Bastian}, {Biermann},
  {Bombrun}, {de Torres}, {Gerlach}, {Geyer}, {Hern{\'a}ndez}, {Hilger},
  {Hobbs}, {Klioner}, {Lammers}, {McMillan}, {Ramos-Lerate},
  {Steidelm{\"u}ller}, {Stephenson}, \& {van Leeuwen}}]{lindegren21}
{Lindegren}, L., {Bastian}, U., {Biermann}, M., {et~al.} 2021, \aap, 649, A4

\bibitem[{{Livio} \& {Soker}(1984)}]{livio1984MNRAS.208..763L}
{Livio}, M. \& {Soker}, N. 1984, \mnras, 208, 763

\bibitem[{{Lodders} {et~al.}(2009){Lodders}, {Palme}, \& {Gail}}]{Lodders2009}
{Lodders}, K., {Palme}, H., \& {Gail}, H.~P. 2009, Landolt B{\"o}rnstein, 4B,
  712

\bibitem[{{Lombardo} {et~al.}(2023){Lombardo}, {Bonifacio}, {Caffau},
  {Fran{\c{c}}ois}, {Jablonka}, {Kordopatis}, {Martin}, {Starkenburg}, {Yuan},
  {Sbordone}, {Sestito}, {Hill}, \& {Venn}}]{lombardo2023}
{Lombardo}, L., {Bonifacio}, P., {Caffau}, E., {et~al.} 2023, \mnras, 522, 4815

\bibitem[{{Lombardo} {et~al.}(2021){Lombardo}, {Fran{\c{c}}ois}, {Bonifacio},
  {Caffau}, {del Mar Matas Pinto}, {Charbonnel}, {Meynet}, {Monaco},
  {Cescutti}, \& {Mucciarelli}}]{lombardo21}
{Lombardo}, L., {Fran{\c{c}}ois}, P., {Bonifacio}, P., {et~al.} 2021, \aap,
  656, A155

\bibitem[{{Luyten} \& {Dartayet}(1942)}]{luyten1942ApJ....96...55L}
{Luyten}, W.~J. \& {Dartayet}, M. 1942, \apj, 96, 55

\bibitem[{{Malaney} \& {Fowler}(1989)}]{malaney1989}
{Malaney}, R.~A. \& {Fowler}, W.~A. 1989, \apjl, 345, L5

\bibitem[{{Mardini} {et~al.}(2022){Mardini}, {Frebel}, {Ezzeddine}, {Chiti},
  {Meiron}, {Ji}, {Placco}, {Roederer}, \&
  {Mel{\'e}ndez}}]{mardini2022MNRAS.517.3993M}
{Mardini}, M.~K., {Frebel}, A., {Ezzeddine}, R., {et~al.} 2022, \mnras, 517,
  3993

\bibitem[{{Mashonkina} {et~al.}(2017){Mashonkina}, {Jablonka}, {Pakhomov},
  {Sitnova}, \& {North}}]{mashonkina17}
{Mashonkina}, L., {Jablonka}, P., {Pakhomov}, Y., {Sitnova}, T., \& {North}, P.
  2017, \aap, 604, A129

\bibitem[{{Mashonkina} {et~al.}(2007){Mashonkina}, {Korn}, \&
  {Przybilla}}]{mashonkina2007}
{Mashonkina}, L., {Korn}, A.~J., \& {Przybilla}, N. 2007, \aap, 461, 261

\bibitem[{{Mayor} \& {Queloz}(1995)}]{mayor1995Natur.378..355M}
{Mayor}, M. \& {Queloz}, D. 1995, \nat, 378, 355

\bibitem[{{Molaro}(2023)}]{molaro23}
{Molaro}, P. 2023, in Memorie della Societa Astronomica Italiana, Vol.~94, 22

\bibitem[{{Molaro} {et~al.}(1997){Molaro}, {Bonifacio}, {Castelli}, \&
  {Pasquini}}]{molaro1997A&A...319..593M}
{Molaro}, P., {Bonifacio}, P., {Castelli}, F., \& {Pasquini}, L. 1997, \aap,
  319, 593

\bibitem[{{Molaro} {et~al.}(2020){Molaro}, {Cescutti}, \&
  {Fu}}]{molaro2020MNRAS.496.2902M}
{Molaro}, P., {Cescutti}, G., \& {Fu}, X. 2020, \mnras, 496, 2902

\bibitem[{{Monaco} {et~al.}(2011){Monaco}, {Villanova}, {Moni Bidin},
  {Carraro}, {Geisler}, {Bonifacio}, {Gonzalez}, {Zoccali}, \&
  {Jilkova}}]{monaco2011A&A...529A..90M}
{Monaco}, L., {Villanova}, S., {Moni Bidin}, C., {et~al.} 2011, \aap, 529, A90

\bibitem[{{Mortier} {et~al.}(2012){Mortier}, {Santos}, {Sozzetti}, {Mayor},
  {Latham}, {Bonfils}, \& {Udry}}]{mortie2012A&A...543A..45M}
{Mortier}, A., {Santos}, N.~C., {Sozzetti}, A., {et~al.} 2012, \aap, 543, A45

\bibitem[{{Myeong} {et~al.}(2019){Myeong}, {Vasiliev}, {Iorio}, {Evans}, \&
  {Belokurov}}]{myeong19}
{Myeong}, G.~C., {Vasiliev}, E., {Iorio}, G., {Evans}, N.~W., \& {Belokurov},
  V. 2019, \mnras, 488, 1235

\bibitem[{{Naidu} {et~al.}(2020){Naidu}, {Conroy}, {Bonaca}, {Johnson}, {Ting},
  {Caldwell}, {Zaritsky}, \& {Cargile}}]{naidu20}
{Naidu}, R.~P., {Conroy}, C., {Bonaca}, A., {et~al.} 2020, \apj, 901, 48

\bibitem[{{Nakamura} \& {Shigeyama}(2004)}]{nakamura2004ApJ...610..888N}
{Nakamura}, K. \& {Shigeyama}, T. 2004, \apj, 610, 888

\bibitem[{{Nielsen} {et~al.}(2023){Nielsen}, {Gent}, {Bergemann}, {Eitner}, \&
  {Johansen}}]{nielsen2023A&A...678A..74N}
{Nielsen}, J., {Gent}, M.~R., {Bergemann}, M., {Eitner}, P., \& {Johansen}, A.
  2023, \aap, 678, A74

\bibitem[{{Nissen} \& {Schuster}(1997)}]{nissen1997A&A...326..751N}
{Nissen}, P.~E. \& {Schuster}, W.~J. 1997, \aap, 326, 751

\bibitem[{{Nomoto} {et~al.}(2013){Nomoto}, {Kobayashi}, \&
  {Tominaga}}]{nomoto2013ARA&A..51..457N}
{Nomoto}, K., {Kobayashi}, C., \& {Tominaga}, N. 2013, \araa, 51, 457

\bibitem[{{Pasquini} {et~al.}(2014){Pasquini}, {Koch}, {Smiljanic},
  {Bonifacio}, \& {Modigliani}}]{pasquini2014A&A...563A...3P}
{Pasquini}, L., {Koch}, A., {Smiljanic}, R., {Bonifacio}, P., \& {Modigliani},
  A. 2014, \aap, 563, A3

\bibitem[{{Plez}(2012)}]{plez2012}
{Plez}, B. 2012, {Turbospectrum: Code for spectral synthesis}, Astrophysics
  Source Code Library, record ascl:1205.004

\bibitem[{{Price-Whelan}(2018)}]{pyia}
{Price-Whelan}, A. 2018, {adrn/pyia: v0.2}

\bibitem[{{Primas} {et~al.}(2000){Primas}, {Molaro}, {Bonifacio}, \&
  {Hill}}]{primas2000A&A...362..666P}
{Primas}, F., {Molaro}, P., {Bonifacio}, P., \& {Hill}, V. 2000, \aap, 362, 666

\bibitem[{{Reeves} {et~al.}(1970){Reeves}, {Fowler}, \& {Hoyle}}]{reeves70}
{Reeves}, H., {Fowler}, W.~A., \& {Hoyle}, F. 1970, \nat, 226, 727

\bibitem[{{Reffert} {et~al.}(2015){Reffert}, {Bergmann}, {Quirrenbach},
  {Trifonov}, \& {K{\"u}nstler}}]{reffert2015}
{Reffert}, S., {Bergmann}, C., {Quirrenbach}, A., {Trifonov}, T., \&
  {K{\"u}nstler}, A. 2015, \aap, 574, A116

\bibitem[{{Ryan} \& {Deliyannis}(1995)}]{ryan1995ApJ...453..819R}
{Ryan}, S.~G. \& {Deliyannis}, C.~P. 1995, \apj, 453, 819

\bibitem[{{Salaris} \& {Cassisi}(2005)}]{salaris05}
{Salaris}, M. \& {Cassisi}, S. 2005, {Evolution of Stars and Stellar
  Populations}

\bibitem[{{Santos} {et~al.}(2002){Santos}, {Garc{\'\i}a L{\'o}pez},
  {Israelian}, {Mayor}, {Rebolo}, {Garc{\'\i}a-Gil}, {P{\'e}rez de Taoro}, \&
  {Randich}}]{santos2002A&A...386.1028S}
{Santos}, N.~C., {Garc{\'\i}a L{\'o}pez}, R.~J., {Israelian}, G., {et~al.}
  2002, \aap, 386, 1028

\bibitem[{{Sbordone} {et~al.}(2014){Sbordone}, {Caffau}, {Bonifacio}, \&
  {Duffau}}]{sbordone14}
{Sbordone}, L., {Caffau}, E., {Bonifacio}, P., \& {Duffau}, S. 2014, \aap, 564,
  A109

\bibitem[{{Schlafly} \& {Finkbeiner}(2011)}]{Schlafly2011}
{Schlafly}, E.~F. \& {Finkbeiner}, D.~P. 2011, \apj, 737, 103

\bibitem[{{Sch{\"o}nrich} {et~al.}(2010){Sch{\"o}nrich}, {Binney}, \&
  {Dehnen}}]{schoenrich10}
{Sch{\"o}nrich}, R., {Binney}, J., \& {Dehnen}, W. 2010, \mnras, 403, 1829

\bibitem[{{Sharma} {et~al.}(2018){Sharma}, {Stello}, {Buder}, {Kos},
  {Bland-Hawthorn}, {Asplund}, {Duong}, {Lin}, {Lind}, {Ness}, {Huber},
  {Zwitter}, {Traven}, {Hon}, {Kafle}, {Khanna}, {Saddon}, {Anguiano}, {Casey},
  {Freeman}, {Martell}, {De Silva}, {Simpson}, {Wittenmyer}, \&
  {Zucker}}]{sharma18}
{Sharma}, S., {Stello}, D., {Buder}, S., {et~al.} 2018, \mnras, 473, 2004

\bibitem[{{Siess} \& {Livio}(1999{\natexlab{a}})}]{siess1999MNRAS.304..925S}
{Siess}, L. \& {Livio}, M. 1999{\natexlab{a}}, \mnras, 304, 925

\bibitem[{{Siess} \& {Livio}(1999{\natexlab{b}})}]{siess1999MNRAS.308.1133S}
{Siess}, L. \& {Livio}, M. 1999{\natexlab{b}}, \mnras, 308, 1133

\bibitem[{{Simpson} {et~al.}(2021){Simpson}, {Martell}, {Buder},
  {Bland-Hawthorn}, {Casey}, {de Silva}, {D'Orazi}, {Freeman}, {Hayden}, {Kos},
  {Lewis}, {Lind}, {Schlesinger}, {Sharma}, {Stello}, {Zucker}, {Zwitter},
  {Asplund}, {da Costa}, {{\v{C}}otar}, {Tepper-Garc{\'\i}a}, {Horner},
  {Nordlander}, {Ting}, {Wyse}, \& {Galah
  Collaboration}}]{simpson2021MNRAS.507...43S}
{Simpson}, J.~D., {Martell}, S.~L., {Buder}, S., {et~al.} 2021, \mnras, 507, 43

\bibitem[{{Sk{\'u}lad{\'o}ttir} {et~al.}(2021){Sk{\'u}lad{\'o}ttir},
  {Salvadori}, {Amarsi}, {Tolstoy}, {Irwin}, {Hill}, {Jablonka}, {Battaglia},
  {Starkenburg}, {Massari}, {Helmi}, \&
  {Posti}}]{skuladottir2021ApJ...915L..30S}
{Sk{\'u}lad{\'o}ttir}, {\'A}., {Salvadori}, S., {Amarsi}, A.~M., {et~al.} 2021,
  \apjl, 915, L30

\bibitem[{{Smiljanic} {et~al.}(2009){Smiljanic}, {Pasquini}, {Bonifacio},
  {Galli}, {Gratton}, {Randich}, \& {Wolff}}]{Smiljanic2009}
{Smiljanic}, R., {Pasquini}, L., {Bonifacio}, P., {et~al.} 2009, \aap, 499, 103

\bibitem[{{Smiljanic} {et~al.}(2008){Smiljanic}, {Pasquini}, {Primas},
  {Mazzali}, {Galli}, \& {Valle}}]{smiljanic2008MNRAS.385L..93S}
{Smiljanic}, R., {Pasquini}, L., {Primas}, F., {et~al.} 2008, \mnras, 385, L93

\bibitem[{{Smiljanic} {et~al.}(2021){Smiljanic}, {Zych}, \&
  {Pasquini}}]{Smiljanic21}
{Smiljanic}, R., {Zych}, M.~G., \& {Pasquini}, L. 2021, \aap, 646, A70

\bibitem[{{Soker}(1998)}]{soker1998AJ....116.1308S}
{Soker}, N. 1998, \aj, 116, 1308

\bibitem[{{Spite} \& {Spite}(1982)}]{Spite1982}
{Spite}, F. \& {Spite}, M. 1982, \aap, 115, 357

\bibitem[{{Steenbock} \& {Holweger}(1984)}]{SH1984}
{Steenbock}, W. \& {Holweger}, H. 1984, \aap, 130, 319

\bibitem[{{Suda} {et~al.}(2008){Suda}, {Katsuta}, {Yamada}, {Suwa}, {Ishizuka},
  {Komiya}, {Sorai}, {Aikawa}, \& {Fujimoto}}]{SAGA}
{Suda}, T., {Katsuta}, Y., {Yamada}, S., {et~al.} 2008, \pasj, 60, 1159

\bibitem[{{Tabernero} {et~al.}(2021){Tabernero}, {Zapatero Osorio}, {Allart},
  {Borsa}, {Casasayas-Barris}, {Demangeon}, {Ehrenreich}, {Lillo-Box}, {Lovis},
  {Pall{\'e}}, {Sousa}, {Rebolo}, {Santos}, {Pepe}, {Cristiani}, {Adibekyan},
  {Allende Prieto}, {Alibert}, {Barros}, {Bouchy}, {Bourrier}, {D'Odorico},
  {Dumusque}, {Faria}, {Figueira}, {G{\'e}nova Santos}, {Gonz{\'a}lez
  Hern{\'a}ndez}, {Hojjatpanah}, {Lo Curto}, {Lavie}, {Martins}, {Martins},
  {Mehner}, {Micela}, {Molaro}, {Nunes}, {Poretti}, {Seidel}, {Sozzetti},
  {Su{\'a}rez Mascare{\~n}o}, {Udry}, {Aliverti}, {Affolter}, {Alves}, {Amate},
  {Avila}, {Bandy}, {Benz}, {Bianco}, {Broeg}, {Cabral}, {Conconi}, {Coelho},
  {Cumani}, {Deiries}, {Dekker}, {Delabre}, {Fragoso}, {Genoni}, {Genolet},
  {Hughes}, {Knudstrup}, {Kerber}, {Landoni}, {Lizon}, {Maire}, {Manescau}, {Di
  Marcantonio}, {M{\'e}gevand}, {Monteiro}, {Monteiro}, {Moschetti}, {Mueller},
  {Modigliani}, {Oggioni}, {Oliveira}, {Pariani}, {Pasquini}, {Rasilla},
  {Redaelli}, {Riva}, {Santana-Tschudi}, {Santin}, {Santos}, {Segovia},
  {Sosnowska}, {Span{\`o}}, {Tenegi}, {Iwert}, {Zanutta}, \&
  {Zerbi}}]{tabernero2021A&A...646A.158T}
{Tabernero}, H.~M., {Zapatero Osorio}, M.~R., {Allart}, R., {et~al.} 2021,
  \aap, 646, A158

\bibitem[{{Tan} {et~al.}(2009){Tan}, {Shi}, \& {Zhao}}]{tan2009}
{Tan}, K.~F., {Shi}, J.~R., \& {Zhao}, G. 2009, \mnras, 392, 205

\bibitem[{{Thornton} {et~al.}(1998){Thornton}, {Gaudlitz}, {Janka}, \&
  {Steinmetz}}]{Thornton98}
{Thornton}, K., {Gaudlitz}, M., {Janka}, H.~T., \& {Steinmetz}, M. 1998, \apj,
  500, 95

\bibitem[{{Tody}(1986)}]{tody86}
{Tody}, D. 1986, in Society of Photo-Optical Instrumentation Engineers (SPIE)
  Conference Series, Vol. 627, Instrumentation in astronomy VI, ed. D.~L.
  {Crawford}, 733

\bibitem[{{Tody}(1993)}]{tody93}
{Tody}, D. 1993, in Astronomical Society of the Pacific Conference Series,
  Vol.~52, Astronomical Data Analysis Software and Systems II, ed. R.~J.
  {Hanisch}, R.~J.~V. {Brissenden}, \& J.~{Barnes}, 173

\bibitem[{{Tonry} \& {Davis}(1979)}]{tonry79}
{Tonry}, J. \& {Davis}, M. 1979, \aj, 84, 1511

\bibitem[{{Umeda} {et~al.}(2002){Umeda}, {Nomoto}, {Tsuru}, \&
  {Matsumoto}}]{Umeda02}
{Umeda}, H., {Nomoto}, K., {Tsuru}, T.~G., \& {Matsumoto}, H. 2002, \apj, 578,
  855

\bibitem[{{Villanova} {et~al.}(2019){Villanova}, {Monaco}, {Geisler},
  {O'Connell}, {Minniti}, {Assmann}, \& {Barb{\'a}}}]{villanova19}
{Villanova}, S., {Monaco}, L., {Geisler}, D., {et~al.} 2019, \apj, 882, 174

\bibitem[{{Woosley} {et~al.}(1999){Woosley}, {Eastman}, \&
  {Schmidt}}]{Woosley99}
{Woosley}, S.~E., {Eastman}, R.~G., \& {Schmidt}, B.~P. 1999, \apj, 516, 788

\bibitem[{{Zacharias} {et~al.}(2013){Zacharias}, {Finch}, {Girard}, {Henden},
  {Bartlett}, {Monet}, \& {Zacharias}}]{ucac4}
{Zacharias}, N., {Finch}, C.~T., {Girard}, T.~M., {et~al.} 2013, \aj, 145, 44

\end{thebibliography}

\end{document}